\documentclass[
 twocolumns
]{aa}  

\usepackage{graphicx}
\usepackage[switch]{lineno}
\usepackage{xcolor}
\usepackage{txfonts}
\usepackage{hyperref}
\hypersetup{
    colorlinks=true,
    linkcolor=blue,
    citecolor=blue,
    urlcolor=blue
}
\begin{document}

   \title{Alfvénic solar wind intervals observed by Solar Orbiter: 
   Exploiting the capability of the SWA plasma suite and source region investigation}

  \author{R. D'Amicis\inst{1}
         \and
         J. M. Raines\inst{2}
         \and
          S. Benella\inst{1}
          \and
         M. Velli\inst{3}
          \and
         O. Panasenco\inst{4}
            \and
         G. Nicolaou\inst{5}
          \and
           C. J. Owen\inst{5}
           \and
          R. M. Dewey\inst{2}
          \and
        P. Louarn\inst{6}
          \and
           A. Fedorov\inst{6}
           \and 
          S. T. Lepri\inst{2}
          \and   
         B. L. Alterman\inst{7}
           \and
          D. Perrone\inst{8}
          \and
           R. De Marco\inst{1}
          \and
           R. Bruno\inst{1}
          \and
          L. Sorriso-Valvo\inst{9,10} 
         \and
          O. S. Dhamane\inst{1}
          \and
          Y. Rivera\inst{11}
          \and
          O. R. Kieokaew\inst{6}
          \and
          D. Verscharen\inst{5}
          \and
            G. Consolini\inst{1}
          \and
          S. Yardley\inst{12}
          \and
          V. Réville\inst{6}
          \and
          D. Telloni\inst{13}
          \and 
          D. Baker\inst{5}
          \and
          G. Lewis\inst{5}
          \and
          G. Watson\inst{5} 
          \and
          C. Anekallu\inst{5}
         \and 
          K. Darwish\inst{5}
          \and
          L. Prech\inst{14}
          \and
          S. Livi\inst{15}
         \and
         T. Horbury\inst{16}
         \and
          G. Mele\inst{17}
          \and
          V. Fortunato\inst{18}
          \and
          F. Monti\inst{19}
      }

   \institute{
   Institute for National Astrophysics (INAF), Institute for Space Astrophysics and Planetology (IAPS), Via del Fosso del Cavaliere, 100, 00133, Rome, Italy\\
   \email{raffaella.damicis@inaf.it}
   \and
    University of Michigan, Ann Arbor, MI, USA
   \and
   University of California Los Angeles, USA
    \and
 Advanced Heliophysics, Pasadena, USA
   \and
  University College London, Mullard Space Science Laboratory, Holmbury St. Mary, Dorking, Surrey, RH5 6NT, UK
   \and
   Institut de Recherche en Astrophysique et Planétologie, CNRS, Université de Toulouse, CNES, Toulouse, France
   \and
   Heliophysics Science Division, NASA Goddard Space Flight Center, Greenbelt, MD, USA 
  \and  
    ASI - Italian Space Agency, Rome, Italy  
   \and
   CNR - National research Council, ISTP - Institute for Plasmas Science and Technology, Bari, Italy
\and
  KTH Royal Institute of Technology, School of Electrical Engineering and Computer Science, Department of Electromagnetics and Plasma Physics, Stockholm, Sweden 
  \and
    Center for Astrophysics, USA
    \and
Department of Mathematics, Physics and Electrical Engineering, Northumbria University, Newcastle Upon Tyne, UK 
\and
Institute for National Astrophysics (INAF) - Osservatorio Astrofisico di Torino, Torino, Italy
 \and
 Charles University, Czech Republich
 \and
  Southwest Research Institute, 6220 Culebra Road, San Antonio TX, 78238, USA
\and
Imperial College London, London, UK
\and
  LEONARDO SpA, Grottaglie, 74023 Taranto, Italy
  \and
  Planetek Italia S.R.L., Via Massaua, 12, 70132 Bari, BA, Italy
  \and
  TSD-Space, Via San Donato 23 - 80126 Napoli, Italy
 }

   \date{Received xx; accepted xx}
   
\titlerunning{Alfvénic solar wind intervals observed by Solar Orbiter SWA}

 
  \abstract
   {
   Fast and slow solar wind have distinct properties linked to their solar sources. The Alfvénic slow wind complicates the usual speed-based classification, especially at intermediate speeds. The Solar Wind Analyzer (SWA) on Solar Orbiter offers unique capabilities to investigate how Alfvénic slow wind differs from classical fast wind and relate these differences to their solar origins.
   }
    {
    In September 2022, Solar Orbiter moved inbound toward the Sun from 0.59 to 0.38 AU, observing several Alfvénic streams: one fast wind, three Alfvénic slow wind (AS1, AS2, AS3), and a moderate fast (FH) interval. We analyze these streams to highlight their similarities and differences.}
   {
We combine plasma parameters from all SWA sensors with magnetic field measurements from the Magnetometer (MAG). A spectral analysis of magnetic and velocity fluctuations is used to characterize Alfvénicity. The magnetic connectivity of each stream to its solar source is examined using Potential Field Source Surface extrapolation combined with ballistic backmapping from the spacecraft.
}
 {
Proton velocity distribution functions show anisotropies and field-aligned beams characteristic of Alfvénic streams, while electron pitch-angle distributions reveal clear strahl populations. Oxygen and carbon charge-state ratios are low in fast wind but higher in AS1–AS3, consistent with slow wind origins; FH shows intermediate values. Magnetic connectivity suggests the fast wind originates from a large coronal hole; AS1 links to a pseudostreamer with high expansion factor; AS2, AS3, and FH connect to a negative-polarity coronal hole whose field lines cross a pseudostreamer that later dissipates. Spectral analysis indicates near energy equipartition in all intervals except AS2.
 }
{
The combined SWA observations offer key insights into the physical processes shaping Alfvénic solar wind streams. Our results reinforce that the simple fast/slow wind classification is inadequate for linking solar wind to sources, and suggest that Alfvénicity relates to the solar source conditions.}


   \maketitle
%
\section{Introduction}

The solar wind is the classical paradigm of a collisionless plasma and a natural laboratory to study fundamental processes in astrophysical plasmas, including turbulence, magnetic reconnection and plasma instabilities \citep[e.g.,][and references therein]{verscharen2019}.
It consists of a continuous flow of particles expanding from the Sun into interplanetary space.
It is mainly composed of protons and electrons, along with a few percent by number of alpha particles (fully ionized helium atoms, $He^{2+}$), and much less abundant heavy ions in different ionization stages (e.g., $O^{6+}$, $C^{5+}$, $Fe^{9+}$).

In situ measurements collected by various spacecraft over almost the last si decades have provided a detailed description of the solar wind over a wide range of distances from the Sun \citep[e.g.,][]{Marsch1982a,Marsch1982b,richardson2004,raouafi2023}, essentially from the corona to the heliopause \citep{Maruca2023,Brown2025}, and also covering a wide range of heliospheric latitudes \citep{bruno1986,schwenn1990,McComas1998,McComas2003,elliott2003} and longitudes \citep{schwenn1990}.
Such observations have shown the need to categorize the solar wind measurements according to different criteria to understand the origin of the wind, as well as its behavior and impacts in the heliosphere.
The traditional classification based on the bulk speed, in which solar wind streams are generally separated into two distinct ``states''---fast and slow---is based on the bimodal distribution of speeds observed near Earth during solar minima \citep[e.g.,][]{Bavassano1991}.
However, this classification is known to be imprecise, especially in the 400 to 600 $\mathrm{km \, s^{-1}}$ range of speeds and during different phases of solar activity \citep{schwenn2006,damicis2011,Fu2018,alterman2025}.

The fast solar wind is typically associated with speeds exceeding 600 $\mathrm{km \,  s^{-1}}$, and is characterized by proton temperatures ($T_p$) that far exceed electron temperatures ($T_e$) in the inner corona, and at least out to 10 $R_S$ from the Sun. At 1 AU, $T_p$ is typically $\sim 2.3$ $\times$ $10^5$ K while $T_e$ is $\sim 1.0$ $\times$ $10^5$ K \citep[e.g,][]{schwenn1990}.
On the other hand, the near-ecliptic slow solar wind has distinct properties: its speed is typically $<$ 500 km s$^{-1}$, and the proton temperature tends to be lower than the electron temperature. The characteristics of the slow solar wind are far more variable than those of the fast solar wind \citep[e.g.,][]{schwenn1990,Marsch1982b}. 

In general, the cause of their different properties can be traced back to their different solar source. While the fast wind originates primarily from coronal holes, the slow solar wind is generally found close to the heliospheric current sheet, originating from closed magnetic field regions that intermittently open into the interplanetary medium, most likely due to reconnection processes, or from high flux tube expansion. However, the sources of the slow solar wind are still a matter of debate \citep[e.g.,][]{WangSheeley1990,noci1997,ofman2000,Antonucci2005,Antiochos,abbo2016}. During the minima of solar activity, the slow wind is usually observed close to the heliospheric current sheet \citep{Smith1978,bav1997} while during maxima, because of the more complex structure of the solar corona \citep{McComas2001}, it is generally associated with the presence of streamers \citep{wang1999} and low-latitude small coronal holes \citep[e.g.,][]{wang1994}.

Fast and slow wind streams differ in terms of the local microphysics that affects the thermodynamic state of all main solar wind species, as well as their relative streaming. 
%
Proton velocity distribution functions (VDFs) in fast streams are characterized by large temperature anisotropies, with the perpendicular temperature component larger than the parallel one ($T_{\perp} > T_\|$) with respect to the magnetic field direction, especially close to the Sun, and by the presence of a secondary proton beam population moving along the magnetic field direction. The proton beam is characterized by a drift velocity relative to the proton core, $V_D$, which is much larger than the Alfvén speed, $V_A$ \citep{feldman1973,Marsch1982b,Goldstein2000,alterman2018}.
On the other hand, in the typical slow wind, the VDFs are nearly isotropic and generally show no significant beams \citep{Marsch1982b,Hellinger2006,matteini2013,alterman2018}.

Alpha particles represent the second most abundant ion population with a number density ratio with respect to protons (i.e. abundance) generally around 4\% in the fast solar wind \citep{neugebauer1962,neugebauer1966,asbridge1974,yermolaev1997,kasper2007,alterman2025a,alterman2025b,alterman2025d}.
This abundance accounts for up to 20\% of the solar wind mass density.
In slow wind, their abundance approaches 1\% or less (depending on the phase of solar activity) \citep{alterman2019,alterman2021,alterman2025c,yogesh2021}.
The VDFs of the alpha particle populations were found to be typically anisotropic with $T_\| > T_{\perp} $ \citep{Marsch1982a}, in contrast to the proton VDFs \citep{Marsch1982b}, although this may be due to the presence of a massive alpha beam \citep[e.g.,][]{brunodemarco} that, historically, has been unresolved due to detector limitations.
In the fast wind, alpha particles are typically 4 or 5 times hotter than protons in the fast wind \citep{Kasper2008,Maruca2013}, or even more \citep{feldman1974,neugebauer1976,feynman1975}. For the slow wind, which is more isothermal, this ratio peaks around $1$ \citep{Kasper2008,Maruca2013,Johnson2023}.
Alpha particle populations  drift with respect to the proton core with speed of the order of the Alfvén speed in the fast wind \citep{Marsch1982a,neugebauer1996,alterman2018} while no differential speed is seen in the typical slow wind \citep{Kasper2008,Maruca2012,alterman2018}. 

Electron VDFs consist of three distinct populations as observed in earlier missions (Helios: \citet{feldman1975,pilipp1987a,pilipp1987b}, Ulysses: \citet{hammond1996,maksimovic1997}, and Wind: \citet{fitzenreiter1998,salem2023}): a thermal core, well described by a Maxwellian distribution with thermal energy $\sim$10 eV; a non-thermal halo, mostly following a $\kappa$-distribution that captures the enhanced high-energy tails, with thermal energy < 80 eV; and a strahl, with bulk energy $\le$ 100 eV.  While the core and halo populations are generally relatively isotropic, the strahl is a strongly field-aligned (or anti-field-aligned) beam of electrons, usually traveling in the anti-Sunward direction that, due to its large drift along the field, that can cause a large anisotropy in the overall VDFs \citep[e.g.,][]{feldman1978,pilipp1987a}. They can also travel both Sunward and anti-Sunward and observations of such bidirectional strahl are thought to be good indicators of newly formed magnetic loops or suprathermal electrons reflected at interplanetary shocks \citep[e.g.,][]{gosling1987,owens2013}. These bidirectional electron heat flux events may be among the more consistent signatures of coronal mass ejections, particularly at 1 au \citep[e.g.,][]{gosling1987}. Understanding unidirectional and bidirectional strahl trajectories, through analysis of electron pitch angle distributions (PADs), can help determine large-scale IMF topology and provide indications of solar connectivity \citep[e.g.,][]{pilipp1987b,owens2013}. 

The strahl likely originates as a more isotropic distribution of electrons in the corona that is focused by the mirror effect as these escaping electrons propagates outward along the weakening magnetic field in the nascent solar wind \citep[e.g.,][]{owens2008}. This may explain the anti-Sunward bulk velocity of the strahl that is usually observed in the solar-wind rest frame.  However, as for the ion beams, a Sunward or bi-directional electron strahl can occur when the magnetic-field configuration changes during the plasma expansion from the Sun \citep[e.g.,][]{gosling1987,owens2017}.
The different electron populations show different properties in the fast and slow wind, with particular reference to temperature anisotropies with respect to the magnetic field direction \citep[e.g.,][]{stverak2008}. This is due to the different collisionality of the two solar wind regimes, with the fast wind being lighter and hotter than the slow one.  \cite{pilipp1987a} illustrated examples of highly collimated, narrow strahl in fast solar wind streams, while the beam was broader, or absent (or fully isotropized) in examples of slower winds.  Moreover, the variation of the strahl width with radial distance has been investigated by a number of authors \citep[e.g.,][]{hammond1996,Anderson_2012,Graham_2017}.  Most recently, \cite{Owen_2022} found that the average pitch angle width of these beams does not show significant evolution with distance, suggesting a balance between focusing and scattering processes across the covered distance range {($\sim0.43$}~-~$\sim1.0~AU$).  However, these data suggest that, on average, the beams broaden with increasing magnetic field strength and narrow  with increasing solar wind speed.  Pitch angle scattering of the strahl has also been suggested as a potential source of the more isotropic halo \citep[e.g.,][]{Vocks_2005,Saito_Gary_2007,verscharen2019b}.
For the sake of completeness, it is worth mentioning the presence of a nearly isotropic superhalo of electrons at energies above $\sim$2 keV \citep{lin1997,lin1998}. They show a power-law energy spectrum of differential flux, J $\sim E^{-\beta}$, with an average index of $\beta \sim$ 2.3.  However, the number density of this population is very small compared to the densities of the other electron species (at 1 au), and its origin remains poorly understood \citep{lin1998,wang2012,yang2015,tao2016}. 




Heavy ions have been shown to be tracers of physical processes in the solar wind and provide crucial information about both the origin and evolutionary processes of the solar wind from the corona into the heliosphere \citep[e.g.,][]{raymond1997,vonsteiger2000,zurbuchen2007}.
Elemental abundance ratios are powerful indicators of solar wind source regions on the Sun \citep{vonsteiger2000}. 
The ionic charge state distributions (the relative abundance of each charge state for a given element compared to the total element abundance) observed in-situ are determined by the temperature, density, and expansion profiles in the inner corona, where a solar wind parcel is originated \citep{geiss1995}.
As a solar wind parcel expands while moving outward from the Sun, these heavy ion charge state distributions are frozen into the values they have in the low corona, so that measurements beyond this point reveal the temperature of the corona at the origin of the parcel.
Charge state ratios are simplified indicators of this, computed from pairs of charge states. 
The most common of these are $O^{7+}/O^{6+}$ and $C^{6+}/C^{5+}$.
The in-situ elemental abundances in the two solar wind regimes also show clear differences: in slow speed wind, heavy elements having low first ionization potential (FIP $\leq$ 10 eV - i.e. low-FIP elements) are typically enhanced by a factor of 3 -- 4 relative to the photosphere, a result of their containment on closed magnetic structures in the corona, while in the fast wind their abundances are more nearly photospheric \citep{geiss1995,vonsteiger2000,zurbuchen1999}. In addition, in the slow solar wind the freeze-in temperature determined from carbon charge states is 1.4 -- 1.6 $\times 10^6$ K while in the fast wind, it is 8 $\times 10^5$ K \citep[e.g.,][]{feldman2005}.

Charge-resolving plasma mass spectrometers, equipped with time-of-flight sections and total energy detectors, are typically able to identify more than ten different heavy elements from carbon to iron and up to 75 individual ion species \citep{gloeckler1998, livi2023}, allowing the study of solar wind sources and their dynamic evolution \citep{vonsteiger2000,zurbuchen2002}.
While the compositional information remains unchanged after the solar
wind escapes the inner corona, the kinetic properties of heavy ions are strongly affected by local processes in the heliosphere \citep{isenberg1983,gary2001}. For example, wave-particle interactions may result in non-thermal distributions of heavy ions. A preferential heating zone, extending to the Alfvén radius, nominally $\sim$15 $R_S$ , has been shown to act both on alpha particles \citep{kasper2017,kasper2019} and heavy ions \citep{holmes2024}. Beyond this zone, Coulomb collisions tend to push particles toward thermal equilibrium in the solar wind rest frame \citep{Maruca2013,Johnson2024}. While heavy ions also transition toward thermal equilibrium, their behavior is more nuanced \citep{tracy2015, tracy2016, holmes2024} and radial evolution may need to be accounted for \citep{Rivera2025b}.



While the differences between fast and slow wind and their mapping to distinct source regions hold in a broad sense, they are known to be imprecise, especially when solar wind composition and turbulence properties are accounted for \citep[e.g.,][]{vonsteiger2008,zhao2009,zhao2017,Stakhiv2015,Xu2015,damicis2015,Camporeale2017,ko2018,Stansby2018,perrone2020,alterman2025,alterman2025a,alterman2025b,alterman2025c,alterman2025d}.
The Alfvénic slow wind \cite[ASW,][]{damicis2011,damicis2015} is an emerging class of solar wind with slow-wind-like speeds, but other properties that are more fast-wind-like.
This plasma is called the ``Alfvénic'' slow wind because the Alfvénic nature of solar wind fluctuations is one of the key quantities for identifying it. 
Alfvénic fluctuations are thought to play a major role in a variety of heliospheric processes. In particular, they are crucial in wave–particle interactions, which ultimately shape non-thermal features—such as anisotropies and beams—observed in ion VDFs and significantly influence ion drifts. 
Given the fundamental impact of Alfvénic fluctuations on plasma dynamics, it is important to recall that Alfvénicity, measured by the presence of correlations between magnetic field and solar wind velocity components and quasi-incompressibility, i.e. almost constant number density and magnetic field magnitude, is a distinguished feature of solar wind fluctuations, as predicted by the ideal magnetohydrodynamics (MHD) theory. It was first observed by Mariner 5 in the ambient fast wind \citep{belcher1969,bd,bs}.
The differences in fast and slow wind Alfvénicity are likely related to the source regions from which they emanate \citep{yardley2024,alterman2025a,alterman2025d}.
The larger the Alfvénicity, the more prominent non-thermal solar wind features typically are and the more the solar wind elemental and charge state composition tends towards values expected for solar wind from coronal holes \citep{damicis2011,damicis2015,damicis2019,damicis2020,stansby2020,perrone2020,damicis2021rev}.
These correlations are typically more reliable than speed, but less well-studied due to their increased complexity.

\citet{alterman2025a,alterman2025d} combine elemental composition and large scale turbulence parameters to create a simplified, unambiguous map for identifying solar wind source regions.
In their map, helium-rich, Alfvénic solar wind likely originates in coronal holes, typically associated with fast wind, and helium-poor solar wind without a significant Alfvénicity is from sources typically associated with slow wind (see, e.g., Figure 11 in \citet{alterman2025a} or Figure 3 in \citet{alterman2025d}).
Under this framework, the maximum speed of plasma from typically slow wind sources is faster than the minimum speed of solar wind from coronal holes due to differences in their in transit acceleration profiles \citep{Halekas2023,Rivera2024,Rivera2025}.
ASW \citep{damicis2011,damicis2015} naturally shows up as the low speed extension of Alfvénic, helium-rich wind.

Alfvénic fluctuations also play a key role in shaping the turbulent nature that characterizes the solar wind \citep[e.g.,][]{brunocarbone2013}.
The turbulent character of the fluctuations is reflected in the power spectra which follow power laws over a wide range of frequencies, although different frequency ranges can be described by different power laws. At low frequencies, an $f^{-1}$ regime, corresponding to fluctuations of the large scales of the turbulent cascade, characterizes the spectrum \citep{matthaeus1986,dmitruk2007,verdini2012,matteini2018,bruno2019a,damicis2020,perrone2020}. The intermediate frequencies, (the so-called inertial range) are typically described by a turbulence spectrum \citep[as first observed by][]{coleman1968}, which is generally described by the Kolmogorov scaling $f^{-5/3}$ \citep{k41}, or by the Iroshnikov-Kraichnan scaling of $f^{-3/2}$ \citep{i63,kraichnan1965}. The power spectra of magnetic field fluctuations show an evolution from a $-3/2$ to a $-5/3$ spectral slope moving away from the Sun \citep[e.g.,][]{Chen2020,Mondal2025}. The power spectra of velocity fluctuations are well described by an Iroshnikov-Kraichnan scaling of $-3/2$ within 1 au \citep{ podesta2006,podesta2007,salem2009,borovsky2012,damicis2020}, and evolve towards $-5/3$ at larger heliocentric distances \citep{roberts2010}. 
Around proton scales, spectra steepen due to kinetic effects \citep{leamon1998,Kiyani2015}. Previous studies \citep{bruno2014,damicis2019} have shown that the spectral index seems to depend on the power characterizing the fluctuations within the inertial range: the higher the power, the steeper the slope.

To better understand the origin and evolution of the Alfvénic slow solar wind, simultaneous observations of the solar sources and in situ measurements close to the Sun, would offer unprecedented insight into its underlying physical processes.
Before the launch of Parker Solar Probe \citep{Fox2016} in 2018, only the observations from the two Helios probes (launched in 1974 and 1976, respectively), allowed the investigation of the radial evolution of relevant solar wind parameters in the ecliptic plane at significant distances inside Earth's orbit, as close as 0.29 au from the Sun. Helios' instrumentation focused on the main solar wind constituents, e.g., electrons, protons and alpha particles.  More recently we have benefited from observations from a number of L1 observatories (e.g., Wind, ACE, DSCVR, Aditya). In 1997, the launch of the ACE spacecraft, equipped with an ion composition spectrometer, provided the first opportunity to determine also the chemical and ionic-charge composition of the solar wind. 
However, only the ESA/NASA Solar Orbiter mission \citep{muller2020}, launched in February 2020, provides us with coordinated remote sensing and in-situ observations, made for the first time on the same spacecraft in the inner heliosphere. 
As we will see in the next section, in situ measurements comprise not only high-resolution plasma instruments able to fully characterize the solar wind but, in particular, a charge-resolving mass composition analyzer that represents an unprecedented opportunity to study heavy ions very close to the Sun. This sensor, along with remote sensing instruments, allows for higher fidelity mapping of the solar wind to its source regions, leading to a better understanding of the solar wind's origin.
 
The Solar Orbiter mission profile can be divided into three phases: the cruise, nominal, and extended phases. After the Earth Gravity Assist Maneuver which occurred on 26 November 2021, Solar Orbiter started its nominal mission, characterized by several close solar encounters at a heliocentric distance of around 0.3 AU. In this paper, we compare different Alfvénic streams observed during this phase of the mission, near the September 2022 perihelion, and identify the different constituents of the solar wind in each stream. We also compare and characterize the fluctuations of the different solar wind intervals to highlight similarities and differences between them.  We investigate their source regions and the properties which may drive these different streams, showing how characterizing the Alfvénicity is a critical for understanding these streams. 

\section{Observations overview}

\begin{figure}
    \centering
    \includegraphics[width=8cm]{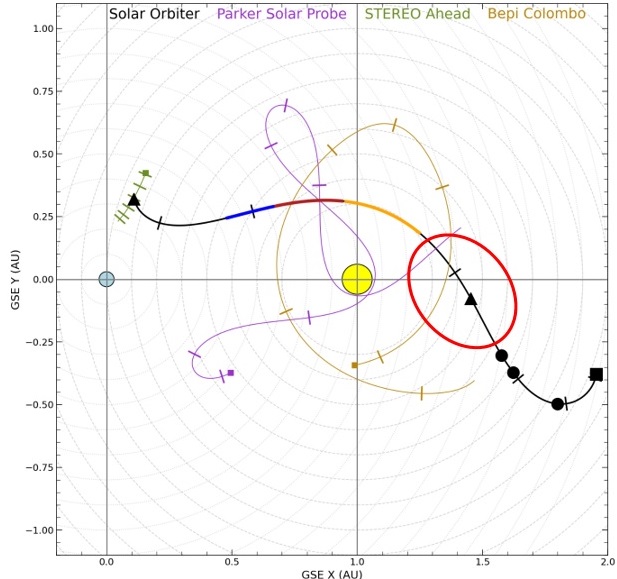}
    \caption{Long Term Planning (LTP) of Solar Orbiter operations 08 \& 09 covering the time interval 2022-06-27 -- 2022-12-26. The figure shows the X-Y projection of the Solar Orbiter orbit (solid black line) in GSE coordinates, along with Parker Solar Probe (purple), Stereo Ahead (green) and Bepi Colombo (yellow). The blue, red and orange segments mark the remote sensing windows of Solar Orbiter. The black square indicates the period start, the triangles the LTP boundaries while the black dots the GAM restrictions. The red oval highlights the interval investigated in this study (adapted from \url{https://2e2.cosmos.esa.int/confluence/display/SOSP/Orbit+Plots})}
    \label{fig00}
\end{figure}

This paper analyzes several Alfvénic intervals observed by Solar Orbiter during the second half of September 2022, preceding the second perihelion passage of the nominal phase of the mission. 
Fig. \ref{fig00} illustrates the Long Term Planning (LTP) of Solar Orbiter operations (LTPs 08 \& 09) covering the time interval 2022-06-27 - 2022-12-26.  A fixed Earth-Sun reference frame is shown, using an X-Y projection of the orbit in GSE coordinates.  The trajectory of Solar Orbiter is plotted as a solid black line.  The figure also shows the trajectories of Parker Solar Probe (purple), STEREO Ahead (green) and Bepi Colombo (yellow). The blue, red, and orange segments along the Solar Orbiter path mark the length of the Solar Orbiter remote sensing windows, during which remote sensing observations are performed. The red oval highlights the interval investigated in the present study. 

\subsection{Instrument details}

The plasma data were provided by the Solar Wind Analyser (SWA)
plasma suite \citep{owen2020}. SWA consists of three sensors dedicated to fully characterize the solar wind: the Proton and Alpha particle Sensor (SWA-PAS), the Electron Analyser System (SWA-EAS) and the Heavy Ion Sensor (SWA-HIS), served by a common Data Processing Unit (SWA-DPU).

SWA-PAS is a top-hat electrostatic analyzer that measures the 3D VDF of the solar wind ions (protons and alpha particles) without mass selection at 92 energies (logarithmically spaced from 200 eV/q to 20 keV/q), 11 azimuth (-24$^{\circ}$ to +42$^{\circ}$) and 9 elevation ($\pm 22.5^{\circ}$) angles (with $\sim6^{\circ} \times 6^{\circ}$ angular resolution). The complete energy/elevation sweeping is performed in 1 s and repeated with a cadence of 1~s or 4~s.
Level 2 (L2) SWA-PAS ground moments (i.e number density, velocity vector, and temperature from the pressure tensor trace) refer to protons only, obtained after removing alpha particles by cutting the 1-D distribution at the saddle point between the two populations. 

SWA-EAS consists of two top-hat electrostatic analyzer heads, mounted orthogonally to each other at the end of a 4-m long boom extending into the shadow of the spacecraft. Each head has a field of view $\sim\,90^{\circ}\,\times\,360^{\circ}$ (elevation $\times$ azimuth).  Due to the orthogonal mounting of the 2 heads, SWA-EAS has a combined field of view that covers the full sky, with the exception of regions blocked by the spacecraft or its appendages. The instrument samples electrons within the energy range from $\sim$1 eV to $\sim$5 keV. The SWA-EAS measurements are used to resolve the 3D VDFs of thermal core, supra-thermal halo and strahl electrons. By combining the 3D VDFs from SWA-EAS with the magnetic field measurements by MAG, and assuming the distributions are gyrotropic, we construct 2D pitch-angle vs energy distribution functions, with one reference axis being along the magnetic field \citep{Owen_2022}. These distributions are convenient to visualize the electron beams, from which we can infer the topological nature of the observed magnetic structure (e.g., switchback, closed field loop, etc.) and its magnetic connection to the solar source.


SWA-HIS is a charge-resolving ion mass spectrometer which uses a Top-Hat electrostatic analyzer paired with a time-of-flight (TOF) telescope. Together they select ions by energy per charge (E/q) before measuring their TOF and total energy (E), allowing separation of solar wind ions by both mass and charge \citep{owen2020}. SWA-HIS scans its full range of E/q, 0.5 keV/q to 75 keV/q, every 30 seconds. These measurements are accumulated over 10 min for publicly released data products. During ground processing, these data pass through a maximum likelihood estimator (MLE) method to assign counts in the TOF-E measurement space to specific ions with minimal uncertainty due to overlap. This MLE method was first developed for ACE/SWICS \citep{shearer2014} and has been adapted for SWA-HIS \citep{livi2023} to derive ion densities, velocities, and thermal speeds. Charge state distributions, and ratios such as $O^{7+}/O^{6+}$, $C^{6+}/C^{5+}$ and $C^{6+}/C^{4+}$, as well as the elemental abundance ratio $Fe/O$ are derived from these SWA-HIS measurements and were used in this study.

Magnetic field measurements are provided by the Magnetometer (MAG) instrument \citep{horbury2020}, averaged at SWA-PAS sampling time. 
Velocity and magnetic field data are given in the heliographic Radial-Tangential-Normal (RTN) coordinate system, where R points away from the Sun toward the spacecraft, T is the cross product of the Sun’s spin axis and R, and N completes the right-handed triad.
Data are available on the Solar Orbiter Archive\footnote{\url{http://soar.esac.esa.int/}}.

\subsection{Event overview and data selection}
 
\begin{figure*}
    \centering
    \includegraphics[width=18cm]{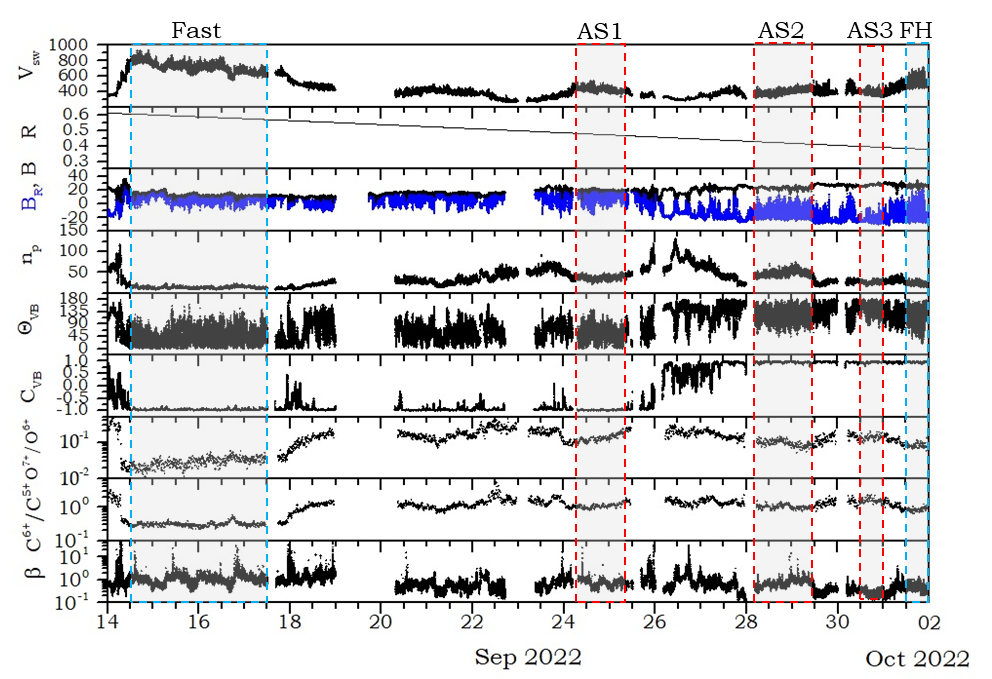}
    \caption{Overview of time series of relevant solar wind parameters observed by Solar Orbiter at a heliocentric distance ranging between 0.58 and 0.32 AU. From top to bottom: solar wind bulk speed, $V_{sw}$, in $km~s^{-1}$; heliocentric distance, $R$, in $AU$; radial component of the magnetic field, $B_R$, in $nT$ (blue) and magnetic field magnitude, $B$, in $nT$ (black); number density, $n_p$, in $cm^{-3}$; the angle the magnetic field forms with the velocity field, $\Theta_{BV}$, in degree; v-b correlation coefficient computed at 30 min scale using a running window, $C_{VB}$; the charge state $O^{7+}/O^{6+}$ ratio; the charge state $C^{6+}/C^{5+}$ ratio; plasma $\beta$. The coloured boxes identify the intervals investigated in this study, corresponding to separate streams of fast wind, Alfvénic slow wind (AS1, AS2, AS3), fast wind (Fast) and moderate fast (FH) solar wind.}
    \label{fig01}
\end{figure*}

Figure \ref{fig01} shows an overview of the whole interval of interest, from 14 September to 2 October, 2022, with the time series of the relevant parameters observed by Solar Orbiter. During this interval, the heliocentric distance of the spacecraft decreased from 0.58 to 0.38 AU. Figure \ref{fig01} shows, from top to bottom: solar wind bulk speed, $V_{sw}$ [$km ~s^{-1}$]; the heliocentric distance, $R$ [$AU$]; the radial component of the magnetic field, $B_R$ (blue), and magnetic field magnitude, $B$ (black), [$nT$]; the proton number density, $n_p$ [$cm^{-3}$]; the angle the magnetic field forms with the velocity vector, $\Theta_{VB}$ [degrees]; the correlation coefficient of velocity and magnetic field components, $C_{VB}$; the charge state $O^{7+}/O^{6+}$ ratio; the charge state $C^{6+}/C^{5+}$ ratio; the plasma $\beta$. 

The parameter $C_{VB}$ is the Pearson's correlation coefficient between V and B components, calculated separately for the three RTN components using a running window of 30 min length \citep[e.g.,][]{tu1989} and then averaged ($\sum_{i=R,T,N} C_{V_iB_i}/3$). The sign of correlation is given by $-sign(V \cdot B)$. Since V is always positive, if the magnetic field has a negative polarity ($B<0$), $V$ and $B$ components are correlated ($C_{VB}$ > 0) while if the magnetic field polarity is positive, V and B components are anticorrelated ($C_{VB}$ < 0).

\begin{table*}
\caption{Bulk parameters averaged over the selected intervals.}             
\label{table:1}      
\centering                          
\begin{tabular}{c c c c c c c c}        
\hline\hline                 
Interval & Begin & End & R & $V_{sw}$ & $n_p$ & $T_p$ & B  \\    
 & & &  [AU] & [$km~s^{-1}$] & [cm$^{-3}$] & [$10^5$ K] & [nT] \\
\hline                        
    F & 2022-09-14T14:24:00 & 2022-09-17T12:00:00 & 0.588 & 704.  & 13.3 & 4.08 & 13.6   \\
    AS1 & 2022-09-24T07:12:00 & 2022-09-25T08:24:00 & 0.474 & 424. & 36.4 & 2.30 & 19.8   \\    
   AS2 & 2022-09-28T04:48:00 & 2022-09-29T10:48:00 & 0.419 & 395. & 48.5 & 2.34 & 23.1   \\
   AS3 & 2022-09-30T12:00:00 & 2022-10-01T00:00:00 & 0.393 & 383. & 25.7 & 2.14 & 27.0   \\
   FH & 2022-10-01T12:00:00 & 2022-10-02T00:00:00 & 0.381 & 520. & 23.2 & 4.09 & 26.6   \\
 \hline                                   
\end{tabular}
\end{table*}

The time interval from September 14 to 18, highlighted in Figure \ref{fig01} by the gray shading labeled `Fast', is dominated by a fast wind stream.
The main portion of this stream is characterized by a bulk speed exceeding 600 $km~s^{-1}$. The kinetic properties of protons and alpha particles of the fast stream were reported in detail by \cite{brunodemarco}.  This fast stream is followed by a rarefaction region, where the bulk speed drops to 400 $km~s^{-1}$, and subsequently by a more standard non-Alfvénic slow wind (around 18 September) with $C_{VB}$ approaching 0. The remaining part of the interval shown in Figure \ref{fig01} is mainly characterized by slow wind streams. 
In general, the $C_{VB}$ coefficient is negative (anti-correlated $V-B$ components) and high, with a well-defined magnetic field (inward) polarity, as in typical Alfvénic intervals. In contrast, just after 26 up to mid 27 September, a fluctuating $C_{VB}$ passing from positive to negative values is observed, indicating the presence of more standard (non-Alfvénic) slow wind, although there are also other intervals where $C_{VB}$ fluctuates and thus identifies non-Alfvénic solar wind (after 14, 18 and 24 September).
The two solar wind regimes identified here are also characterized by different amplitudes in velocity and magnetic field fluctuations, which are seen to be larger in Alfvénic slow streams. In this particular case, only the $B_R$ component is shown, but it is representative to show the amplitude of the fluctuations in the different streams. We must also take into account the dependence of Alfvénicity on radial distance. Indeed, in this interval, the heliocentric distance decreases with time, which may determine the larger amplitude of the fluctuations observed in the Alfvénic slow wind compared to the earlier fast wind.  
The analysis which follows in this paper is based on the comparison of a number of selected intervals identified by the shaded regions between the colored vertical dashed lines in Figure \ref{fig01}: `AS1', `AS2' and `AS3' identify portions of Alfvénic slow wind with similar bulk speed (average speed 400 km/s), `Fast' or `F' is a fast wind interval and `FH' is an intermediate wind (`FH' stands for `fastish', say moderate fast). Table \ref{table:1} shows the average bulk parameters for the selected intervals.  The Alfvénic slow wind intervals are identified on the basis of high $C_{VB}$ values, large-amplitude and weakly-compressible fluctuations (indicated by almost constant magnetic field and number density).  $C_{VB}$ is however chosen only for the preliminary selection of the intervals.  The full Alfvénic range of every stream will be investigated in Section \ref{spectral} by performing a spectral analysis.

During the interval of interest, we observe two heliospheric current sheet crossings, one at the beginning of the interval and the other one just before Sep 26. This allows us to identify two magnetic field polarities for the observed solar wind streams. Indeed, the F and AS1 intervals are characterized by the same magnetic field (outward) polarity while AS2, AS3 and FH by an opposite (inward) polarity. 
All intervals contain evidence for the presence of switchbacks, namely large and intermittent polarity reversals of the radial component of the magnetic field, (see Fig. \ref{fig01}) with amplitudes comparable to the magnitude of the magnetic field \citep{borovsky2016,Bale2019}. It should be noted that not only does the 
$B_R$ component change sign within these structures, but the other components can also vary \citep{DudokDeWit20}. 
The occurrence and properties of switchback signatures change with heliocentric distance \citep[e.g.,][]{Jagarlamudi2023}.


\section{Proton and alpha VDFs observed by SWA-PAS}

\begin{figure*}
    \centering
    \includegraphics[width=18cm]{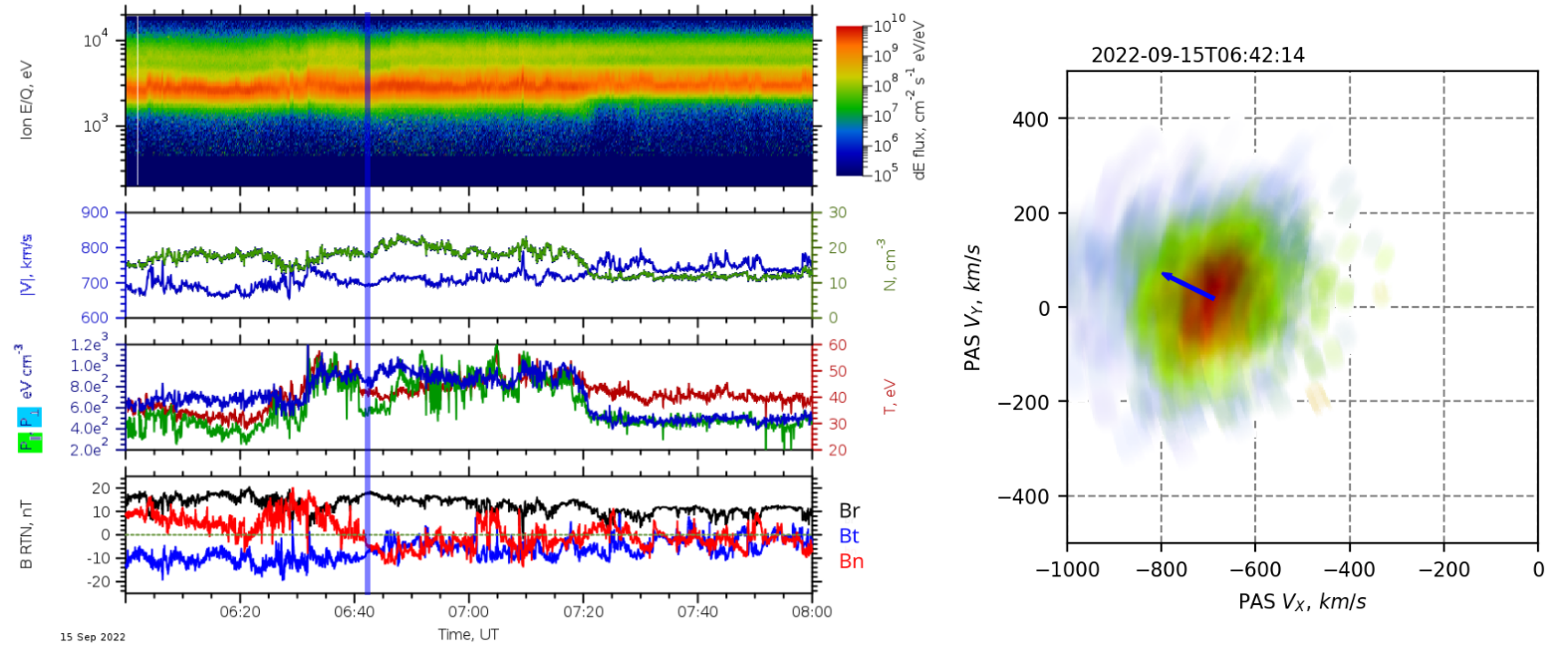}
    \caption{SWA-PAS and MAG observations during the "Fast" solar wind interval.  
\emph{Left panel (top to bottom):} Energy–time spectrogram of the ion energy differential flux;  
solar wind velocity (blue) and number density (green);  
parallel pressure $P_{\parallel}$ (green), perpendicular pressure $P_{\perp}$ (blue), and temperature $T$ (red);  
and magnetic field vector components in RTN coordinates.  
The vertical blue line indicates the time of the VDF shown in the right panel.  
\emph{Right panel:} Ion velocity distribution function (VDF) in the $\mathbf{V} \times \mathbf{B}$ plane.  
The X-axis is aligned as closely as possible with the spacecraft–Sun direction. The blue arrow shows the magnetic field vector.}
    \label{ET_spectrogram_fast}
\end{figure*}

\begin{figure*}
    \centering
    \includegraphics[width=18cm]{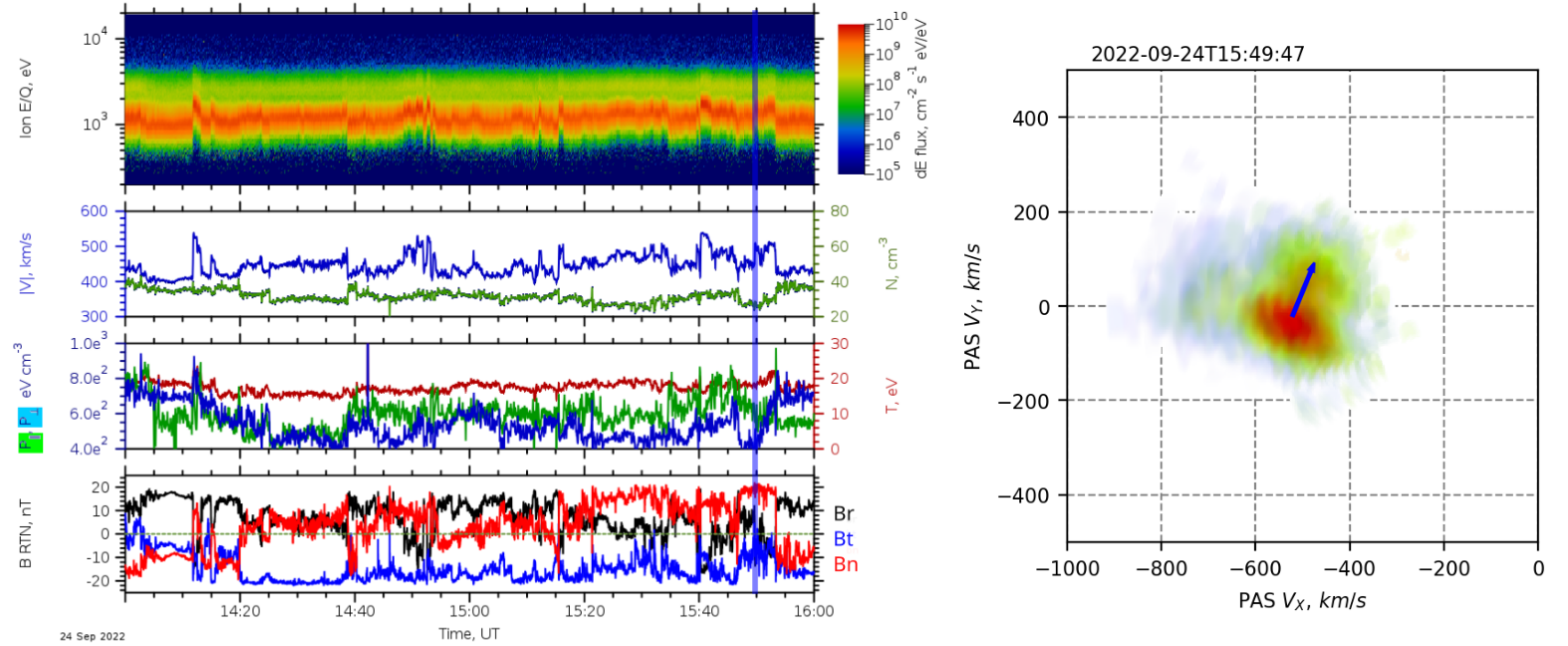}
    \caption{The same as \autoref{ET_spectrogram_fast}, but for the slow solar wind case, corresponding to the time interval AS1 (see  \autoref{fig00} ).}
    \label{ET_spectrogram_slow}
\end{figure*}

SWA-PAS and MAG observations performed in the fast solar wind interval (September 15, 06:00–08:00)  
and the AS1 interval (September 24, 14:00–16:00 UT) are presented in \autoref{ET_spectrogram_fast} and \autoref{ET_spectrogram_slow}, respectively.  
In both cases, an animated sequence showing the time evolution of the ion velocity distribution function (VDF) in the $\mathbf{V} \times \mathbf{B}$ plane is provided in the supplementary materials. The VDF images presented in these figures were generated using the "Direct Volume Rendering" technique, commonly applied in computed tomography.  The 3D VDF is treated as a transparent object, where the internal optical density and color are functions of the VDF value.  
Light rays are cast through the volume from behind, and color and opacity are accumulated along each ray to form the final 2D image.  
For further details, see \citet{Callahan2008}.

For the fast wind (\autoref{ET_spectrogram_fast}), we have selected, as an example, a time period during which the flow shows a transition between a perturbed state and a more stable one. In the energy spectrogram, the main proton population is observed at $\sim$ 2.5-4.5 keV and the alpha particles appear as a yellow/green band, centered at 7-8 keV. Before 0720 UT, the flow is perturbed, with density, velocity, temperature and pressure fluctuations of $\sim 10-20 \%$ on time scales of a few 10 s. In particular, the pressure anisotropy ($P_{\perp} /P_{\parallel}$) abruptly varies from 1 to 2 and vice versa over a few second time scales. After 0720 UT, the flow is comparatively more stable. It slightly accelerates (from $\sim$ 700 to 750 $km~s^{-1}$), decreases in density ($\sim$ 20 to 12 cm$^{-3}$) and becomes almost isotropic ($P_{\perp} \sim P_{\parallel} $). 

For the slow Alfvénic wind (\autoref{ET_spectrogram_slow}), the selected period is representative of the characteristic high level of turbulence of the flow. A key aspect is the almost perfect correlation between the magnetic and velocity fluctuations, with $\mathbf{\delta B}$ and $\mathbf{\delta V}$ related by the Alfvén wave relation (see also \autoref{spectral}). The magnetic field is constant in magnitude, so that its fluctuations are simply fast rotations that almost perfectly correlate with strong variations (up to $\sim$ 100 $km~s^{-1}$) seen in any component of the velocity.  

\begin{figure*}
    \centering
    \includegraphics[width=18cm]{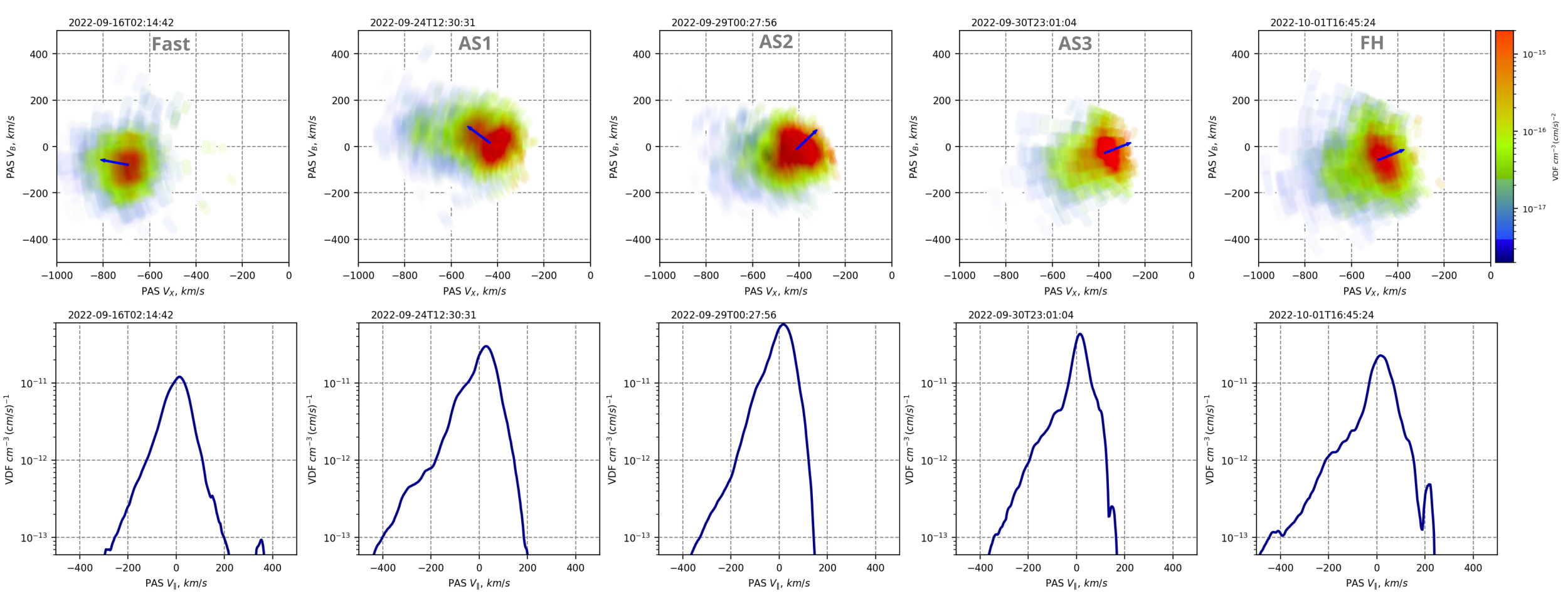}
    \caption{SWA-PAS 2D (upper panel) and 1D (lower panel) ion velocity distribution functions (VDFs) for the selected intervals (see \autoref{fig01}).  
The 2D distributions are integrated onto the plane perpendicular to the $\mathbf{V} \times \mathbf{B}$ vector and include the solar wind bulk velocity vector.  The bottom row shows the $V_{\parallel}$ cut of each VDF.}
    \label{charact_VDFs}
\end{figure*}

SWA-PAS characteristic ion velocity distribution functions (VDFs) corresponding to each type of solar wind shown in \autoref{fig01} are presented in \autoref{charact_VDFs}. In particular, 2D distributions integrated in the plane perpendicular to the $\mathbf{V} \times \mathbf{B}$ vector and including the solar wind bulk velocity vector are shown in the upper panels. In the bottom row, the $V_{\parallel}$ cut of each VDF is displayed.
The Fast 2D VFD displays a typical anisotropic distribution, characterized by $P_{\perp} \ge P_{\parallel}$.  
These VDFs can be described as classical bi-Maxwellian distributions. Their dominant anisotropic core appears as the red oval.  
A small secondary population (proton beam) is also observed along the magnetic field direction. The core and proton beam populations are shown in more detail in the $V_{\parallel}$ profiles.  
The proton beam is located approximately 150–250 $km~s^{-1}$ from the main peak, which is about twice the local Alfvén speed ($\sim$100 $km~s^{-1}$). This is in agreement with \cite{brunodemarco} who studied the kinetic features of the different ion populations of the same fast wind. Moreover, a similar result was found by \cite{damicis2025} for another fast wind interval with Solar Orbiter data. In any case, this result is in agreement with previous studies using data from past missions  \citep{feldman1973,Marsch1982b,marschlivi1987,Goldstein2000}. For this particular case, the proton beam represents less than 5\% of the total ion population. This estimate is again in agreement with \cite{brunodemarco} and with previous studies \citep{neugebauer1962,neugebauer1966,asbridge1974,yermolaev1997,kasper2007}.
In the cases presented here, since the magnetic field is not aligned with the bulk flow, alpha particles are only marginally captured in the direction parallel to \textbf{$B$}. The secondary population observed in all $V_{\parallel}$ profiles of \autoref{fig01} is predominantly composed of protons.

The FH (intermediate solar wind, almost fast wind) VDFs, although anisotropic, differ from classical bi-Maxwellians due to their extended tails aligned with \textbf{$B$}. 
Their $V_{\parallel}$ profiles deviate from a Gaussian shape, instead appearing triangular when plotted on a logarithmic scale.  
This is a characteristic feature of so-called ‘hyperbolic’ distributions.
As discussed in \citet{louarn_24}, this type of proton distribution, for which it is difficult to clearly distinguish a beam from the core, is not uncommon. \citet{louarn_24} proposed to directly fit them by a skewed distribution, as the Normal Inverse Gaussian (NIG), introduced by \citet{barndorff_97} and now widely applied in various fields, such as finance, turbulence, biology, and technical modeling. 
%
%


All AS distributions show a well-developed beam component, corresponding to nearly 30\% of the main proton peak. The beam is located at $\sim $ 120 $km~s^{-1}$ from the core population, which is about 1.6 the Alfvén velocity in the present case. This is in agreement with a recent study by \cite{damicis2025} that compared the drift speed of the proton beam of different Alfvénic slow wind intervals. In many cases, the gaussian fit of the core distribution appears excellent. However, as in the fast wind, there are also situations for which the core/beam distinction is more questionable. An example is given with AS2. Compared to VDF AS1 and AS3, the beam has been eroded and the phase space density within the gap between the core and the beam has been filled. 

The reader is invited to watch the animation in the supplement material to observe the respective dynamics of the core and beam linked to the passing of the strong Alfvénic fluctuations. In particular, it is interesting to note that the core of the distribution is apparently more dynamic than the beam, which is comparatively more stable in the RTN frame.





\section{Investigation of Alfvénicity}
\label{spectral}

\begin{figure}
    \centering
    \includegraphics[width=7cm]{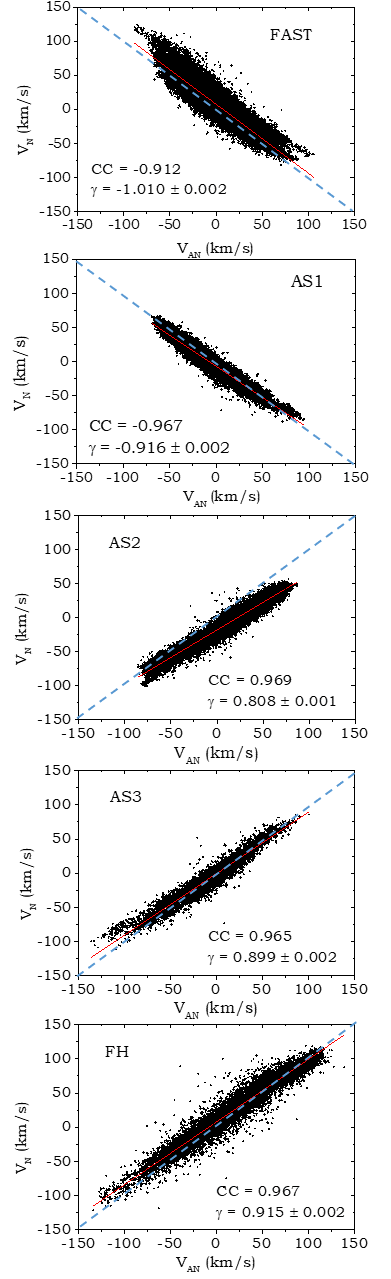}
    \caption{Scatter plots of the normal components of velocity, $V_N$ vs. the normal component of the magnetic field, $V_{AN}$, in Alfvén units. The Pearson correlation coefficient, CC, along with the slope of the fit, $\gamma$, represented by the solid red line, is shown in each panel. The blue dashed line, $V_N = - V_{AN}$ or $V_N = V_{AN}$, indicating equipartition of energy, are shown for comparison.}
    \label{fig:corr}
\end{figure}

Alfvénic fluctuations can be identified by the presence of correlations between velocity and magnetic field components. In order to select the most Alfvénic part of the fluctuations, we should mainly focus on the $N-T$ plane, since the radial component $R$ is the least Alfvénic. In this section, however, we will consider only the normal component $N$ (see Figure \ref{fig:corr}), which is more Alfvénic than the other two components \citep{tu1989}.  
In ideal conditions, $\mathbf{\delta V} = \pm \mathbf{\delta b}$, where $\mathbf{\delta b}$ is the magnetic field fluctuating component given in Alfvén units (i.e. $\mathbf{\delta b}=\mathbf{\delta B}/\sqrt{\mu_0 \rho}$, with $\mu_0$ and $\rho$ the magnetic permeability and mass density, respectively). 
The mass density, in L2 data, refers only to the contribution provided by protons. If we represent one component of $\delta V$ vs $\delta b$ in a scatter plot, we should in principle expect to find a linear relationship with a slope equal to 1. However, this is not what observations usually show \citep[e.g.,][]{bav2000}. The slope of the linear fit provides information on the energy balance of the fluctuations. Indeed, the slope, $\gamma$, of the fit is equivalent to the square root of the Alfvén ratio  $r_A = e^V/e^B$, where $e^V$ and $e^B$ are the kinetic energy and magnetic energy, respectively. From the fit, we can derive also the Pearson's correlation coefficient, CC.
Derived Alfvén ratios, $r_A$, are the following: $1.02$, $0.839$, $0.653$, $0.808$, $0.837$ moving from the upper to the lower panel of Figure \ref{fig:corr}. The ideal slope equal to $\pm1$, indicating equipartition of energy, is shown as a blue dashed line while the red solid line is the linear fit. This ideal condition is encountered in the fast wind. 
In general, all other streams, except for AS2 (showing a large magnetic energy imbalance), are also quite close to equipartition of energy. 

\begin{figure}
    \centering
    \includegraphics[width=\columnwidth]{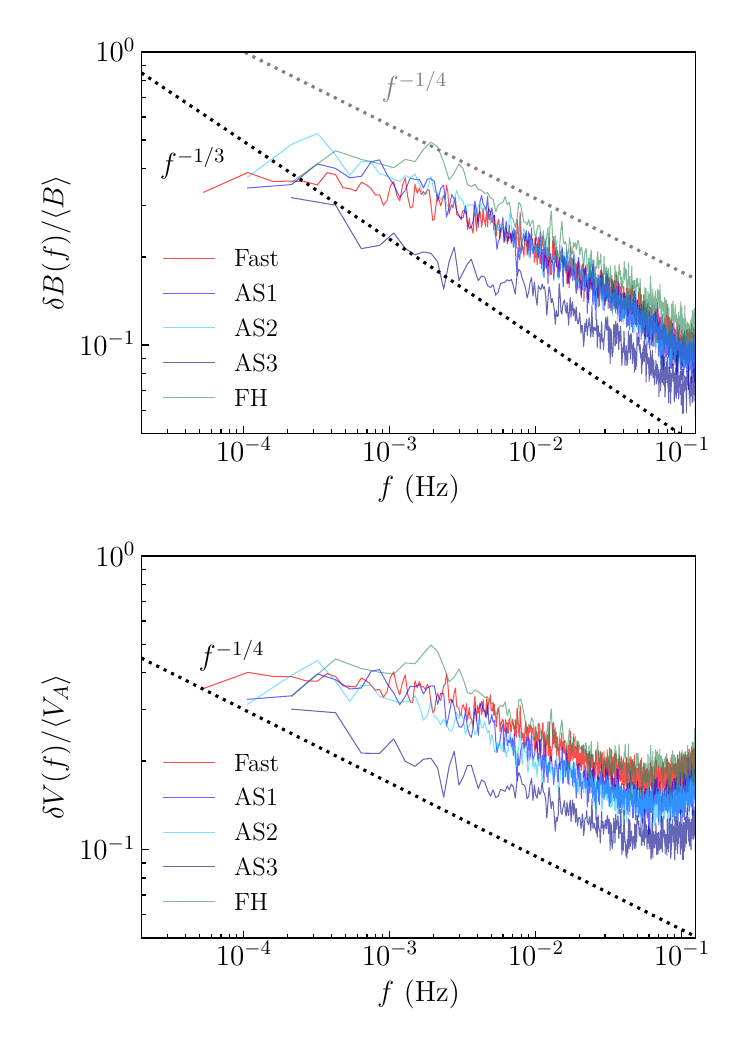}
    \caption{Normalized PSD of the trace of magnetic field and solar wind velocity, upper and lower panels respectively, for Fast (red), AS1 (dark blue), AS2 (light blue) and AS3 (blue) and FH (green) data intervals. Dashed lines indicate the typical slopes of the IK ($f^{-1/4}$) and K41 ($f^{-1/3}$) scalings.}
    \label{fig06b}
\end{figure}

Although the scatter plots shown in Figure \ref{fig:corr} give a quick look indication of the degree of correlations characterizing the different streams and the energy balance of the fluctuations, they do not allow to study the whole range of scales characterizing Alfvénic turbulence. In this context, magnetic and velocity fluctuations can be fully characterized performing a spectral analysis.
The power spectra of the trace of magnetic field and solar wind velocity will be investigated in a follow-up paper, with particular reference on the evolution of the power associated to solar wind fluctuations, of the slopes of the different frequency ranges and spectral breaks separating them as a function of the heliocentric distance.
In this paper, instead, normalized power spectra of the trace of $B$ and $V$ are shown to compare the different streams avoiding the effect of the radial evolution ($\Delta R = 0.2$ AU) that determines a different amplitude of the fluctuations in different streams. The normalized power spectra are computed in the following way. We derived the relative amplitude of the fluctuations from the Fourier analysis to compute the trace of magnetic field as $\delta B(f)/\langle B \rangle = \sqrt{2 f S_B(f)}/\langle B \rangle$, where $S_B(f)$ is the trace of the PSD of the magnetic field components. In the same way, we derived $\delta V(f)/\langle V_A \rangle$.
Normalizing to $\langle B \rangle$ and $\langle V_A \rangle$ allows us to compare the fluctuations of different streams \citep{bruno2019a,damicis2020,damicis2022}.
The normalized power spectra ($\delta B(f)/\langle B \rangle$ and $\delta V(f)/\langle V_A \rangle$) in the upper and lower panels, respectively, are shown in Fig. \ref{fig06b}. According to this normalization, Kolmogorov and Kraichnan scalings can be easily derived. Indeed, since $S_B(f)\sim f^{-5/3}$, $\delta B(f)/\langle B \rangle \sim f^{-1/3}$ for the Kolmogorov scaling. However, in the lower-frequency MHD range $S_B(f)\sim f^{-3/2}$ or, equivalently, $\delta B(f)/\langle B \rangle \sim f^{-1/4}$, closer to the Kraichnan spectrum. The observation of two different sub-ranges within the inertial domain is in agreement with recent observations in the inner and outer heliosphere \citep{Mondal2025,Wu2025}, as well as previous Solar Orbiter observations \citep{damicis2025}. In a similar way, $\delta V(f)/\langle V_A \rangle \sim f^{-1/4}$ (since $S_V(f)\sim f^{-3/2}$) for the Kraichnan scaling.

The v-b correlation coefficient reported in Fig. \ref{fig01}, used as a  proxy to identify and select Alfvénic solar wind intervals, has been computed at a fixed typical Alfvénic scale of 30 minutes. 
Similar to \cite{marschtu1990}, in this section, we perform a spectral analysis using standard tools to study turbulence in the frequency domain: the normalized cross-helicity, $\sigma_C$, and the normalized residual energy, $\sigma_R$.
As described in Appendix \ref{A}, $\sigma_C$ is derived from the Els\"asser variables \citep{elsasser} to investigate the Alfvénic content of solar wind fluctuations and the full range of frequencies where Alfvénicity is high \citep{tu1989,grappin}.

Since Alfvénic turbulence is characterized by high v-b correlations along with weak compressibility, meaning that fluctuations in the magnitude of the magnetic field are much smaller than the fluctuations in the components of the magnetic field, magnetic compressibility is shown as well. This basically shows a fundamental property of Alfvénic fluctuations that they tend to stay in a state of spherical polarization \citep{bruno2001, matteini2015}. Magnetic compressibility, $C_B$, is defined as the ratio between the PSD of the magnitude of magnetic field over the trace of PSD of the magnetic field components. 

\begin{figure*}
    \centering
    \includegraphics[width=18cm]{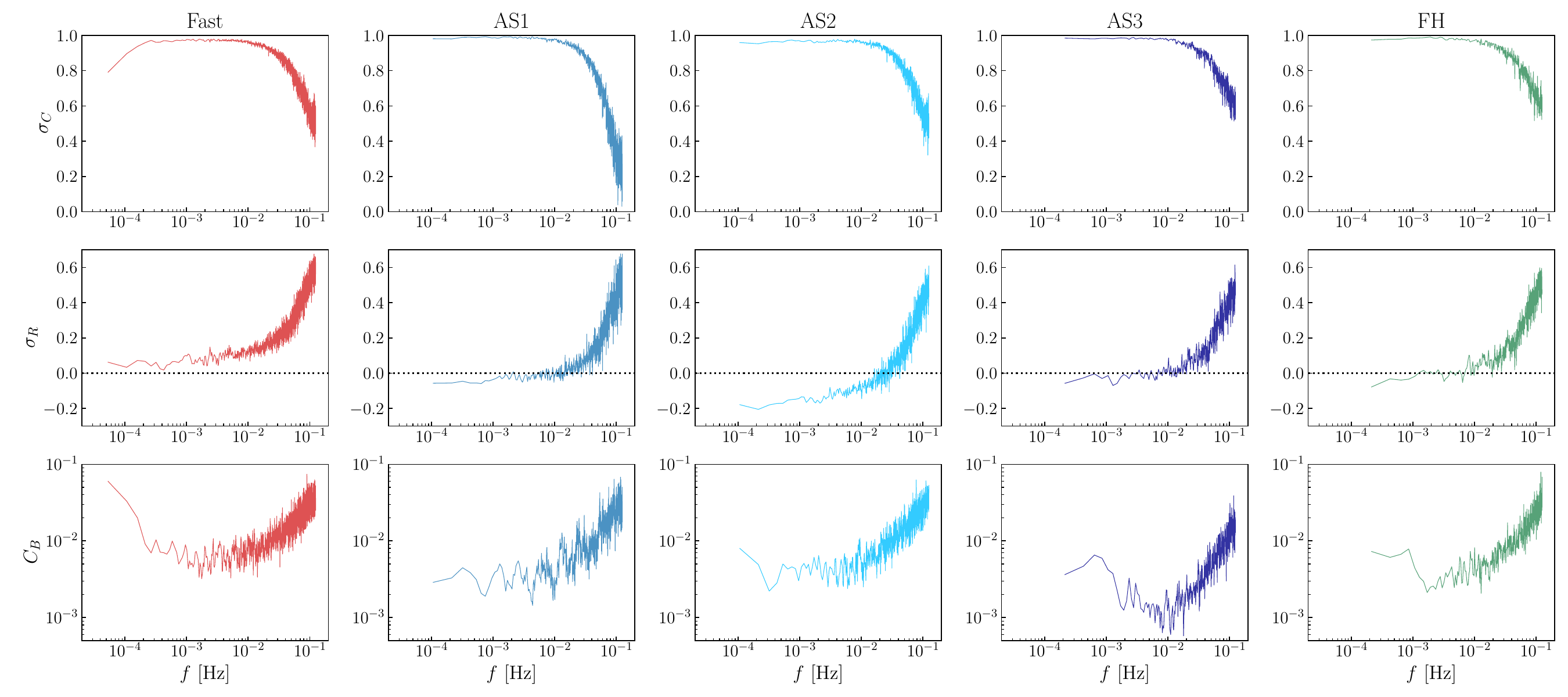}
    \caption{Power Spectra of the normalized cross-helicity, $\sigma_C$ (upper), the normalized residual energy, $\sigma_R$ (middle) and magnetic compressibility $C_B$ (lower) for the fast wind (red), for the different Alfvénic slow wind streams, AS1-AS3 (different shades of blue) and for the FH (green) solar wind interval.}
    \label{fig07}
\end{figure*}

Figure \ref{fig07} shows the power spectra of $\sigma_C$ (upper panels), $\sigma_R$ (middle panels) and $C_B$ (bottom panels) of the different streams (same color code used throughout the paper). 
The comparison between the different streams shows a similar extension of the Alfvénic range that extends up to 4 $\times 10^{-2}$ Hz, in agreement with \cite{damicis2025}. 
In general, all intervals show similar and high $\sigma_C$ values in the range $2 \times 10^{-4} - 10^{-2}$ Hz, confirming previous observations by Helios \citep[e.g.,][]{tumarsch1995} and recent findings by PSP \citep[e.g.,][]{Thepthong2024,damicis2025b}. On the other hand, $\sigma_R$ appears to be close to 0 in the frequency range $10^{-3}-10^{-2}$ Hz for AS1, AS3 and FH and slightly negative for lower frequencies. While $\sigma_R$ is always negative for AS2, highlighting that magnetic energy dominates kinetic energy, it shows a tendency to have positive values in the whole frequency range in the fast wind. This apparent kinetic energy imbalance in the fast wind is related to the neglect of alpha particles, as shown in Appendix \ref{B}.
The results of the present analysis are not consistent with previous papers. Indeed, the fast wind was found to be closer to equipartition of energy than the Alfvénic slow wind both at L1 \citep{damicis2022} and near 0.35 au \citep{damicis2025}, with the Alfvénic slow wind showing a magnetic energy imbalance even close to perihelion. This would suggest that equipartition of energy is, in general, lost very rapidly probably due to some local compressive phenomena \citep{damicis2025}. 


\section{Electron VDFs measured by SWA-EAS}
\begin{figure*}
    \centering
    \includegraphics[width=18cm]{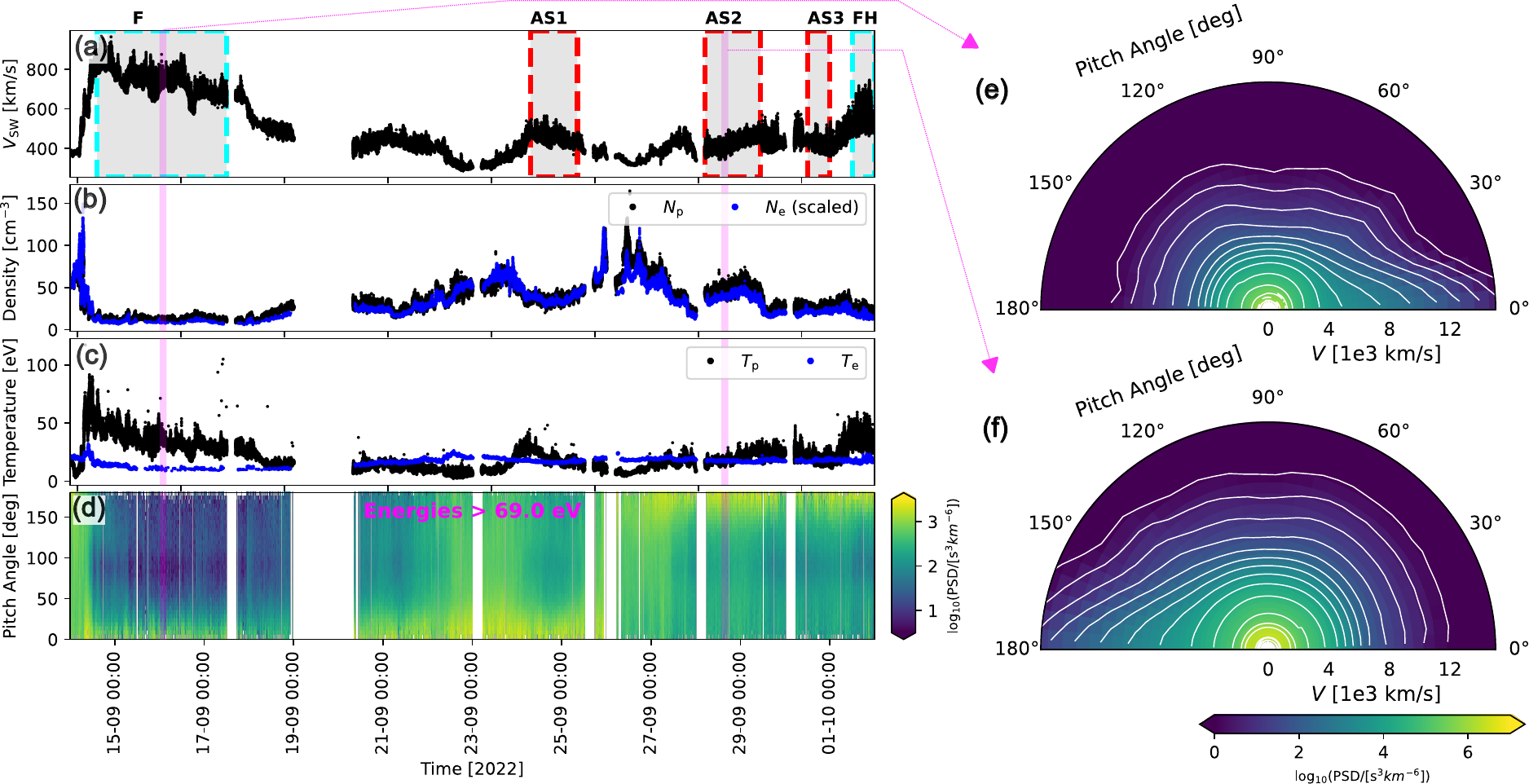}
    \caption{ Time series of \textbf{(a)} proton speed, \textbf{(b)} electron (blue) and proton (black) densities, \textbf{(c)} electron (blue) and proton (black) temperatures, \textbf{(d)} pitch-angle distribution of "strahl" electrons for energies >69 eV, \textbf{(e)} a 2D gyrotropic, pitch-angle vs energy distribution averaged over a sub-interval in the middle of the fast wind interval, and \textbf{(f)} a  distribution for a sub-interval within the AS2 interval.} 
    \label{fig03}
\end{figure*}

 Figure \ref{fig03}\textbf{(a)} shows the time series of the plasma proton speed for the period from 14 September to 02 October.  The shaded regions again mark the selected intervals of interest (similar to the top panel in Figure \ref{fig01}). The blue traces in panels \textbf{(b)} and \textbf{(c)} show time series of the plasma "core" electron density $N_{\mathrm{e}}$ and temperature $T_{\mathrm{e}}$, respectively, along with the corresponding protons parameters (black traces). These electron "core" parameters are determined by fitting an 1D, non-drifting Maxwellian distribution function model to the thermal energy range of the observed electron phase space density (PSD). The energy range < 12 eV is excluded from this fitting analysis, since below this energy we observe mostly secondary- and photo-electrons emitted by the spacecraft \citep{nicolaou2021,stverak2025}. Moreover, the PSD we use for the fitting analysis is corrected for the spacecraft potential \citep{lavraudlarson2016}. We also scale the determined $N_{\mathrm{e}}$ with a factor (constant for the entire interval) that accounts for uncertainties in the absolute gain of the sensors, and brings the plotted values to be consistent with the $N_{\mathrm{p}}$ levels. Although this is based on the assumption that $N_{\mathrm{e}}$ $\sim$ $N_{\mathrm{p}}$, for our purposes, however, it is more important that we find a very similar variation of $N_{\mathrm{e}}$ and $N_{\mathrm{p}}$, which validates the accuracy of the determined parameters. Interestingly, the temperature of the two species does not show a correlation in their variation. The plasma electrons behave much more isothermally across the entire interval. For the fast solar wind intervals, $T_{\mathrm{e}}$ is often smaller that $T_{\mathrm{p}}$. In the slow wind intervals ($V_{\mathrm{SW}} < 400  \mathrm{km\,s^{-1}}$), however, $T_{\mathrm{p}}$ exceeds $T_{\mathrm{e}}$. 
 
We further construct the 2D pitch-angle vs energy distribution function of the solar wind electrons, taking into account the average magnetic field direction during the acquisition of each SWA-EAS measurement sample. Figure \ref{fig03}\textbf{(d)} shows the time series of the pitch angle distribution of electrons with energies > $\sim69$ eV. At this energy range we clearly observe the field-aligned electron beams ("strahl"). Although there is variability of the "strahl" pitch angle distribution, we observe that for the selected intervals, there is a rather distinct and persistent PSD peak along the parallel ($0^{\circ}$), or anti-parallel ($180^{\circ}$) directions. We also observe a change of the beam direction, from parallel to anti-parallel, at the beginning of Sept 26, 2022. This change occurs across the reversal of the radial magnetic field component, along with an enhancement of proton density, consistent with a current sheet crossing. In Figure \ref{fig03}\textbf{(e)} and \textbf{(f)}, we show examples of the full pitch-angle vs energy (speed) distribution averaged over 2 sub-intervals (one in the fast stream marked 'F' and the other in AS2) that are indicated by the two vertical magenta ribbons. Both examples show a clear beam signature. In the first case the beam is peaked along the magnetic field direction, while in the second case it is flowing in the anti-parallel direction. 

\begin{figure*}
    \centering
    \includegraphics[width=12cm]{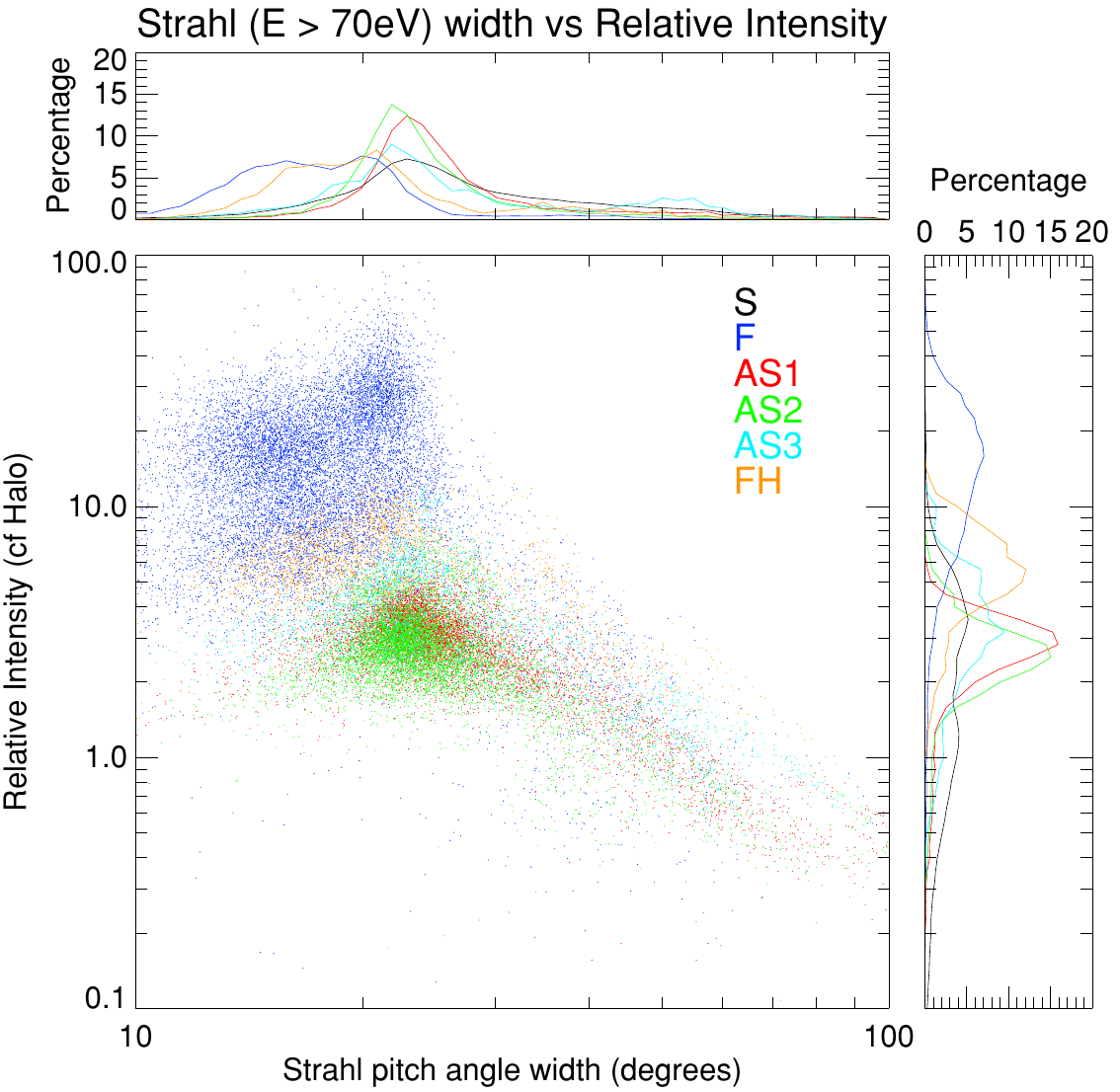}    \caption{Main panel:  Scatter plot of results of fitting analysis to the SWA-EAS strahl electron (E > $\sim69$ eV) pitch angle distribution.  The horizontal axis indicates the fitted pitch angle width of an individual sample, with the vertical axis representing the relative intensity of the sampled strahl beam with respect to the concurrent halo population. Upper panel:  Histograms of the percentage of samples falling within one of 50 logarithmically spaced pitch angle bins in range ($10-100^{\circ}$).  Right hand Panel:  Histograms of the percentage of samples falling within one of 60 logarithmically spaced relative intensity bins in range 0.1-100.  In each of the 3 panels, the different regions identified within the overall period of interest are represented by different colors: Blue 'F'; Red 'AS1'; Green 'AS2', Teal 'AS3'; Orange 'FH'. All points that fall outside the periods of interest are fitted as black 'S'. In the top and right-hand panels, the histogram for the remaining samples in this period are shown, although these are surpressed from the main scatterplot panel for clarity.}
    \label{fig_strahl}
\end{figure*}

We provide an animation in the supplementary material that shows the evolution of the 2D pitch-angle vs energy distribution throughout the entire interval. An examination of the electron PAD's across the entire interval, as presented in this animation, will show that in numerous cases within the (non-Alfvénic) slow solar wind (i.e. many periods outside the selected intervals marked in Figures \ref{fig01} and \ref{fig03}), the electron beams do not stand-out as a prominent feature as much as they do during the highlighted sub-intervals. We seek to quantify these statements by fitting the strahl population observed throughout this entire interval using the method described in \cite{Owen_2022} and examining the differences in strahl width and relative intensity with respect to the concurrent halo population.  These parameters are derived from fits to the pitch angle distribution for E > $\sim$70 eV, with the latter derived from the relative intensity of the near field-aligned (or anti-aligned) flux compared to that observed for particles moving near-perpendicular to the field. A scatter plot of our results is shown in the main panel of Figure \ref{fig_strahl}. In this plot we have identified the individual fits within each of the regions identified above by a colored point: Blue points - 'F' (19903 fitted samples); Red points - 'AS1' (7362 samples); Green points - 'AS2' (8956 samples); Teal points - 'AS3' (3131 samples); Orange points - 'FH' (3548 samples).  There are also 57699 samples which fall outside of the identified regions during the period of interest (black 'S'), but these are not included in the scatter-plot for reasons of clarity. They are however included in the sub-plots above and to the right of the main panel, which show the percentage of a given sample set falling within column-summed and row-summed data bins respectively.  The upper panel shows percentages of measured points falling in each of 50 logarithmically spaced bins in the range $10-100^{\circ}$), while the right panel shows the percentage of samples falling within one of 60 logarithmically spaced relative intensity bins in range 0.1-100.  The same color scheme as used in the main panel is deployed in these subplots, with the result of the fits from periods outside those specifically identified shown as the black trace. From Figure \ref{fig_strahl}, it is clear that the 'F' and 'FH' intervals typically exhibit narrower and more intense beams than the other identified ('AS') regions, or indeed the background slower solar wind. The most typical beam width for the 3 'AS' regions is similar, and similar also to the slow wind observations, although their typical intensity may be slightly lower than in the slow wind where a strahl population is identified (relative intensity values near 1 in this analysis suggest the distribution is near isotropic in this energy range, such that no field aligned beam stands out from the halo population).  
  
The comparison here between the fast and non-Alfvénic slow wind is consistent with the results of \cite{pilipp1987a} and \cite{Owen_2022} who report more highly collimated, narrow strahl in fast solar wind streams, but broader, or even fully isotropized distributions at strahl energies in slower winds. Although collimated beams appear in both fast and Alfvénic slow wind streams in our interval of interest, the latter are typically broader and less intense than in the former stream. This may indicate that the Alfvénic slow wind streams may have experienced more intense and/or longer duration scattering processes on their journey to the measurement point. The electron populations within the highlighted sub-intervals thus may typically carry more heat-flux when compared to those within the non-alfvenic streams within the examined interval. 

On the other hand, there is no significant difference between the core electron temperature $T_{e}$ during the highlighted intervals and $T_{e}$ during the slow, non-Alfvénic wind time periods. In addition, the $T_{e}$ variations are significantly smaller than the observed $T_{p}$ variations within the entire interval. However, since $T_{e}$ is calculated using the preliminary calibration factors and assuming a non-drifting isotropic Maxwellian core, further study is needed to properly quantify the electron bulk parameters. 




\section{Solar wind composition measured by SWA-HIS}

We used the composition information in Figure \ref{fig01} to support the identification of the intervals under study. Values in both the $O^{7+}/O^{6+}$ and $C^{6+}/C^{5+}$ charge state ratio panels are lower in the identified 'Fast' interval than during most of the rest of the time of interest, as typical of fast or coronal hole related wind \citep{vonsteiger2011}. In contrast, during the AS1, AS2, AS3 intervals, the O and C ratios are clearly higher, reflecting values that are more similar to what would be expected for the slow, or non-coronal hole related solar wind \citep{geiss1995}. While the actual O ratio during AS1, AS2, AS3 may not be above the 0.145 slow solar wind threshold reported in \citet{zhao2009}, this still reflects a more likely slow wind associated composition during these time intervals.  \citet{lepri2013} showed that composition ratios evolve over the solar cycle in both the slow and fast wind, and that, along with any differences in instrumentation characteristics between ACE/SWICS and the Heavy Ion Sensor, may contribute to uncertainties in the absolute values of the ratios. The FH interval has C and O ratios that lie between slow wind values and fast wind values.
Since O and C ratio are believed to freeze-in at similar heights in the low corona \citep{buergi1986,chen2003,landi2012}, these ratios have been observed to be generally well correlated in long-base line studies, with the exception of so-called outlier wind \citep{zhao2017}.  In the case of the outlier solar wind, the C ratio is observed to break the correlation with the O ratio and has much lower values than expected.  Occurring primarily in slow solar wind and at the higher O ratio values, these low C ratio values are thought to be associated with a depletion in the density of fully stripped $C^{6+}$ \citep[see][]{zhao2017}.   

\begin{figure}
    \centering
    \includegraphics[width=9cm]{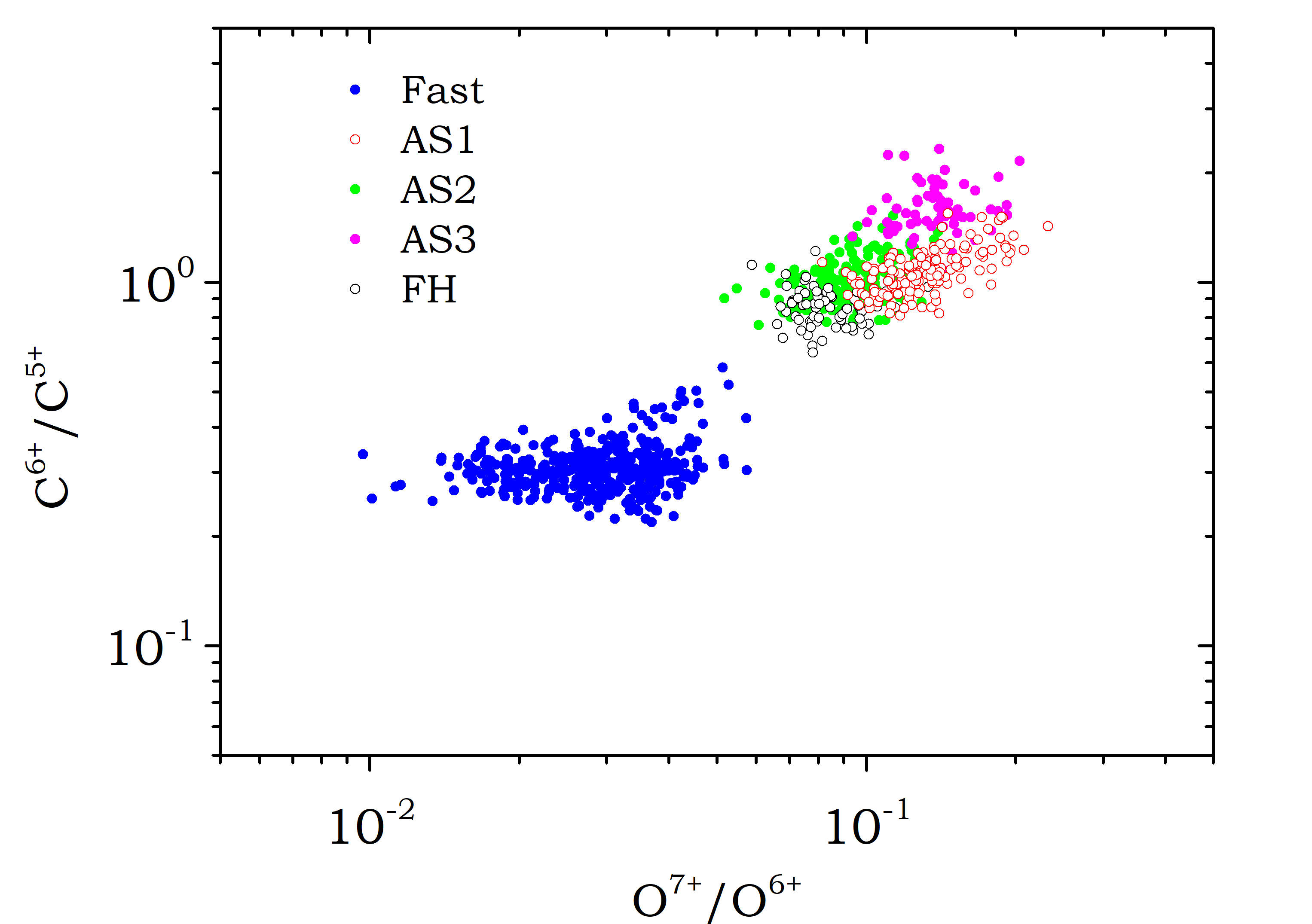}
    \caption{Scatter plot of oxygen ratio ($O^{7+}/O^{6+}$) vs. carbon ratio ($C^{6+}/C^{5+}$) for the selected intervals.} 
    \label{fig04b}
\end{figure}

An examination of the C and O ratio correlation for all five intervals is shown in Figure \ref{fig04b}.  The F interval (blue) is clearly separated from the other intervals AS1-3 and FH, which are partially overlapping one another in this space. Furthermore, it can be seen from the correlation of these clusters of C and O ratios that these intervals do not fall into the ranges typical of outlier wind which extend to lower values of $C^{6+}/C^{5+}$, while maintaining higher values of $O^{7+}/O^{6+}$.  The FH interval appears quite close in composition to the AS intervals, though its speed is closer to that of the FAST interval, which is why it received a separate designation of FH. The separation between FAST and AS intervals is in contrast to a previous study using ACE data from 2002 \citep{damicis2019} where the fast and slow Alfvénic intervals were clustered together.  The slow wind in that study had an O ratio closer to 1.0 and a C ratio near 5, and the fast and Alfvénic wind were clustered near an O ratio of 0.1 and a C ratio near 0.7. That study examined intervals during maximum of solar activity (2002) of solar cycle 23.  In the current study, which was observed during the ascending phase of solar cycle 25, the slow Alfvénic wind and the FH wind look much like the fast and Alfvénic wind from 2002. However, the F wind is very different, appearing at much lower O and C ratios than the 2002 wind.  Based on the findings of \citet{lepri2013}, the O and C ratio in the fast solar wind are about a factor or 3-5 lower during solar minimum than during solar maximum. This appears to indicate that the much lower fast wind O and C ratios observed during our 2022 interval is more similar to solar minimum fast wind (coronal hole wind), while the 2002 fast wind is more similar to solar maximum fast wind.  Based on the \citet{damicis2019} paper and this current analysis, the Alfvénic slow wind occupies a composition regime in between the fast and slow solar wind.   These results agree with recent work by \citet{ervin2024}, who showed composition characteristics of highly Alfvénic slow wind that was traced to an origin near an over-expanded coronal hole boundary. They also agree with \citet{alterman2025}, who show that slow-intermediate speed solar wind is likely from coronal holes, but is not as 'Alfvén wave rich' as what is traditionally identified as fast wind due to differences in the source region topology.

\section{Connecting to the solar sources}

In the previous sections, we have thoroughly examined the in situ characteristics of the solar wind. We now turn our attention to investigating the sources of the solar wind, aiming to understand the origin and physical mechanisms driving its variability.

\begin{figure*}
    \centering
    \includegraphics[width=12cm]{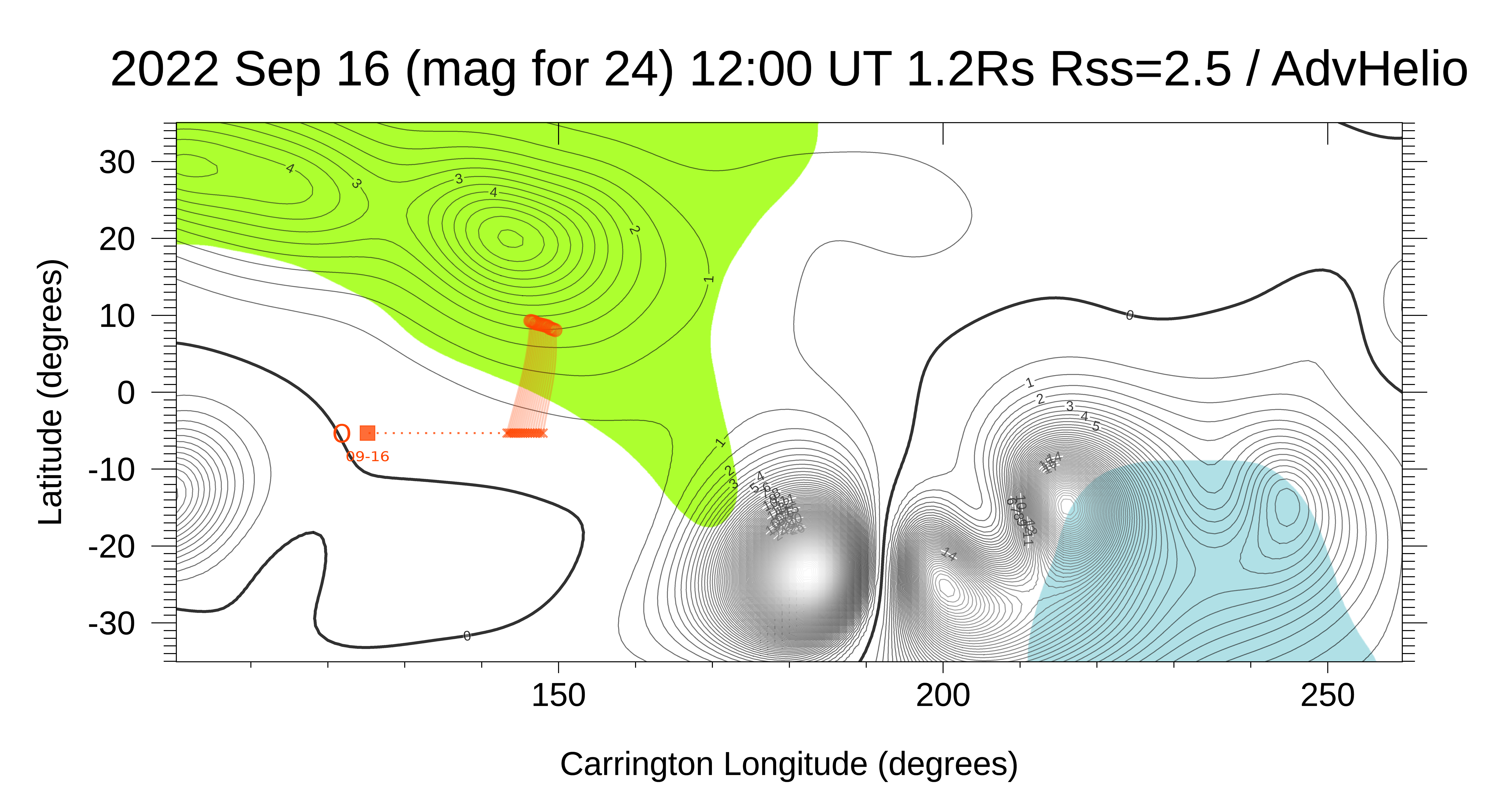}
    \includegraphics[width=7cm]{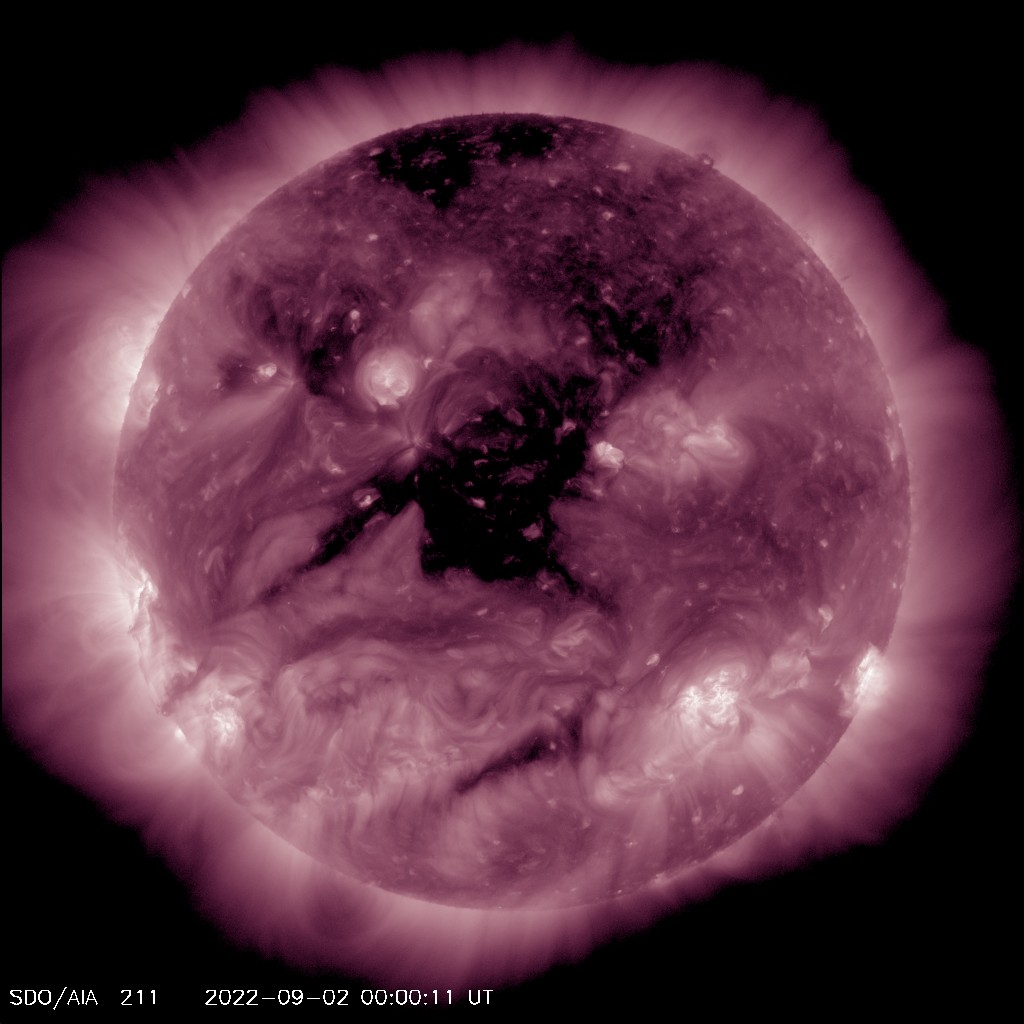}
     \caption{Upper panel: Potential field source surface $B^2$ contour maps and solar wind magnetic foot-points along Solar Orbiter  trajectories (red) for the first Fast interval. The map shows magnetic pressure iso-contours calculated for the heights R = 1.2 R$_{\odot}$ and the projection of the s/c location (squares) on the source surface (crosses) and down to the solar wind source region (circles). The crosses result from ballistic mapping using the measured in situ solar wind speed $\pm80$ $km~s^{-1}$ in bins of 10 $km~s^{-1}$. Open magnetic field regions are shown in green (positive polarity) and cyan (negative polarity) while the neutral line is in black bold. In this period, Solar Orbiter was connected to the large coronal hole with wide southward extension. Lower plot: Image from SDO/AIA of the 211$\AA$ band showing the corona and source region for the interval. The image date, reported in the panels, was chosen to best display the source region on the Sun.}
    \label{fig05a}
\end{figure*}

\begin{figure*}
    \centering
   \includegraphics[width=12cm]{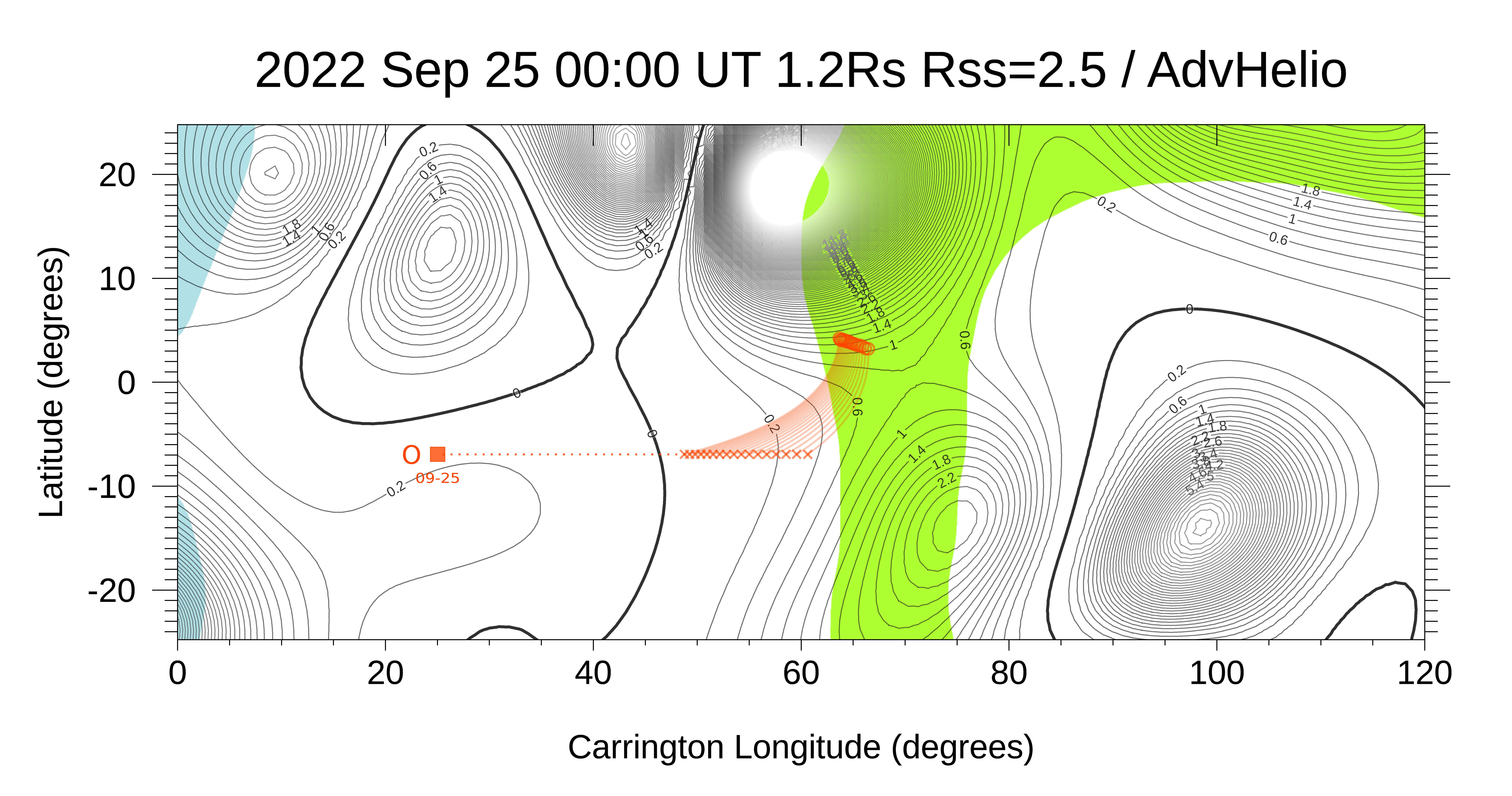}
  \includegraphics[height=7cm]{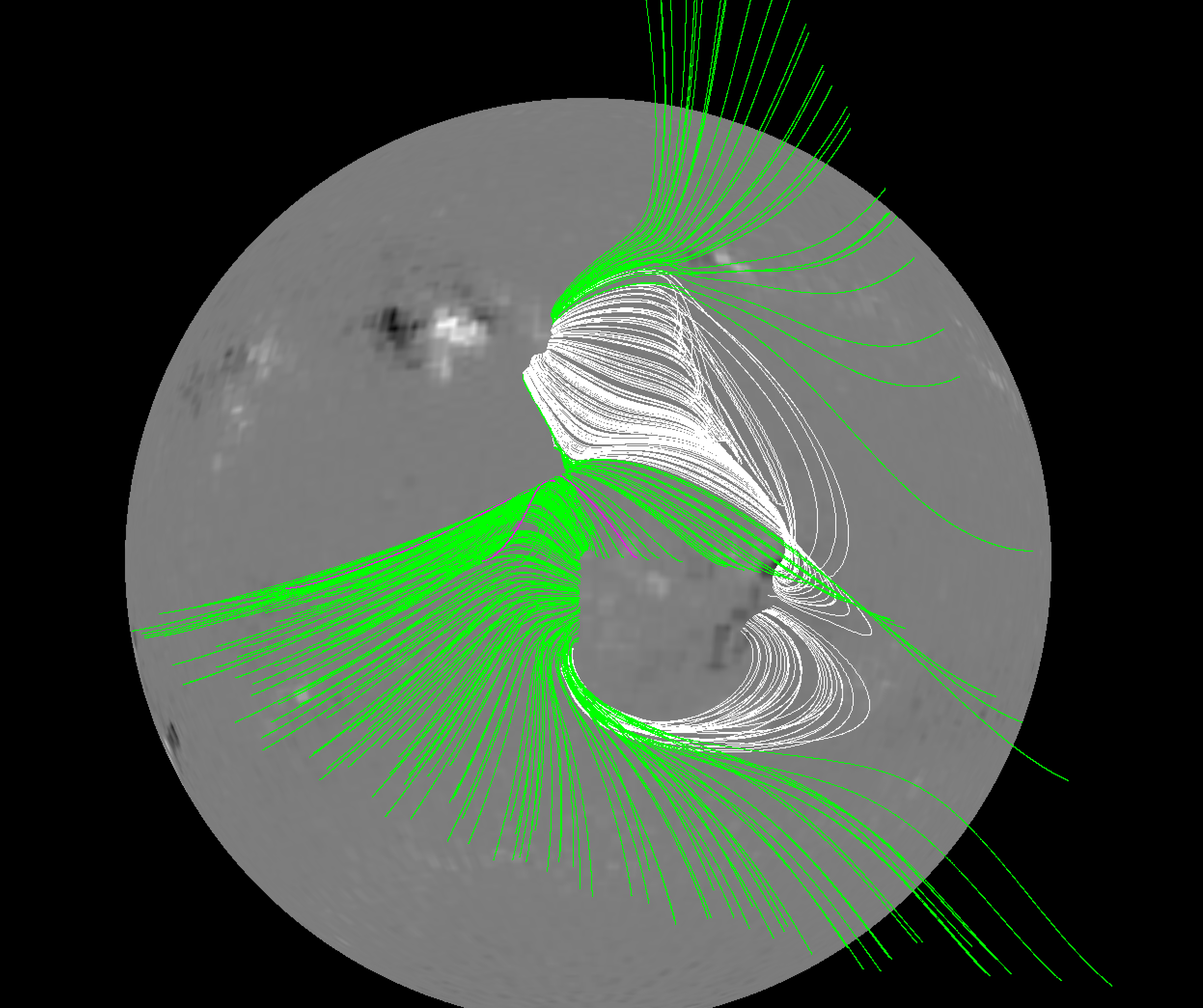} \includegraphics[height=7cm]{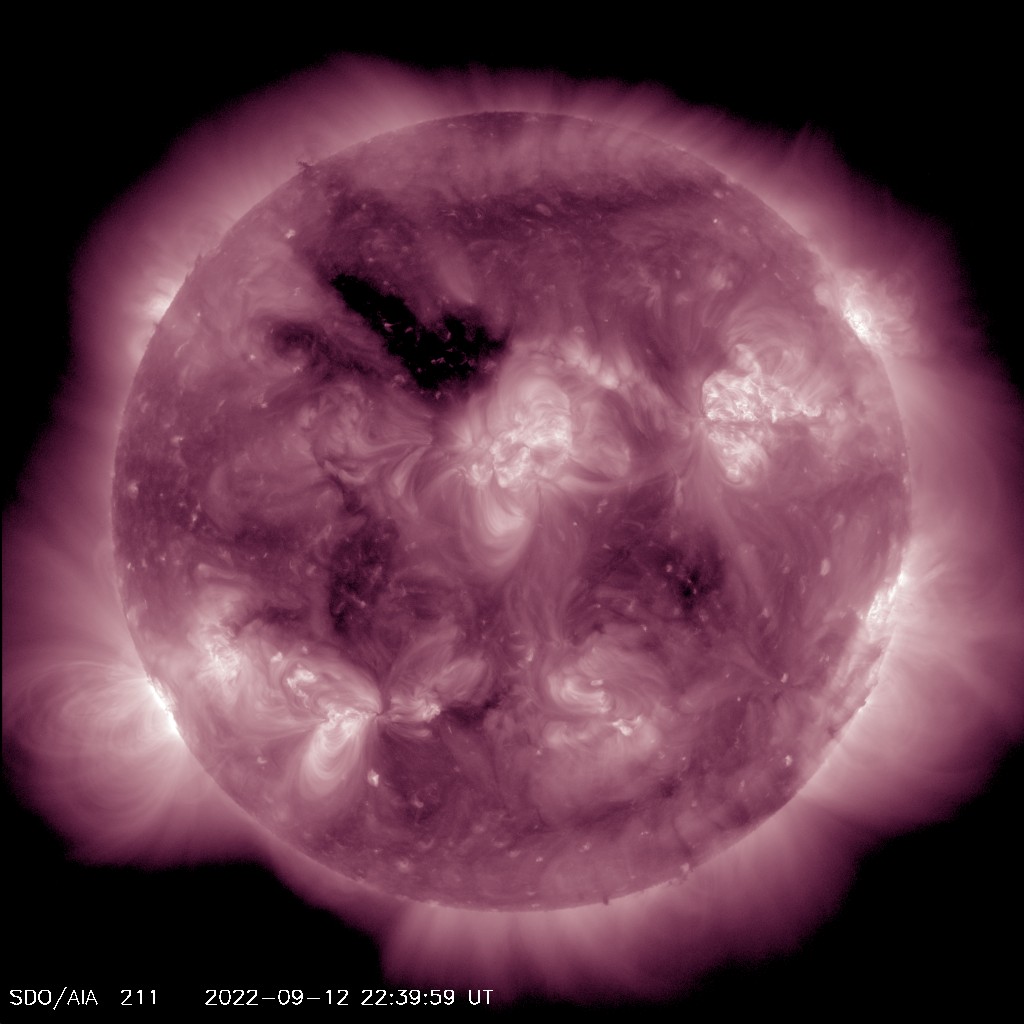}
       \caption{Upper panel: Potential field source surface $B^2$ contour maps and solar wind magnetic foot-points along Solar Orbiter  trajectories (red) for the selected intervals. The map shows magnetic pressure iso-contours calculated for the heights R = 1.2 R$_{\odot}$ and The projection of the s/c location (squares) on the source surface (crosses) and down to the solar wind source region (circles). The crosses result from ballistic mapping using the measured in situ solar wind speed $\pm80$ $km~s^{-1}$ in bins of 10 $km~s^{-1}$. Open magnetic field regions are shown in green (positive polarity) and cyan (negative polarity) while the neutral line is in black bold. Middle panel: 3D potential field source surface model of the pseudostreamer for AS1 interval. The modeling made along the area represented in the upper panel. Lower panel: Image from SDO/AIA of the 211$\AA$ band showing the corona and source regions of the  solar wind stream AS1 described in the paper. The image date, reported in the panel, was chosen to best display the source region on the Sun.}
    \label{fig05b}
\end{figure*}

\begin{figure*}
    \centering
    \includegraphics[width=11cm]{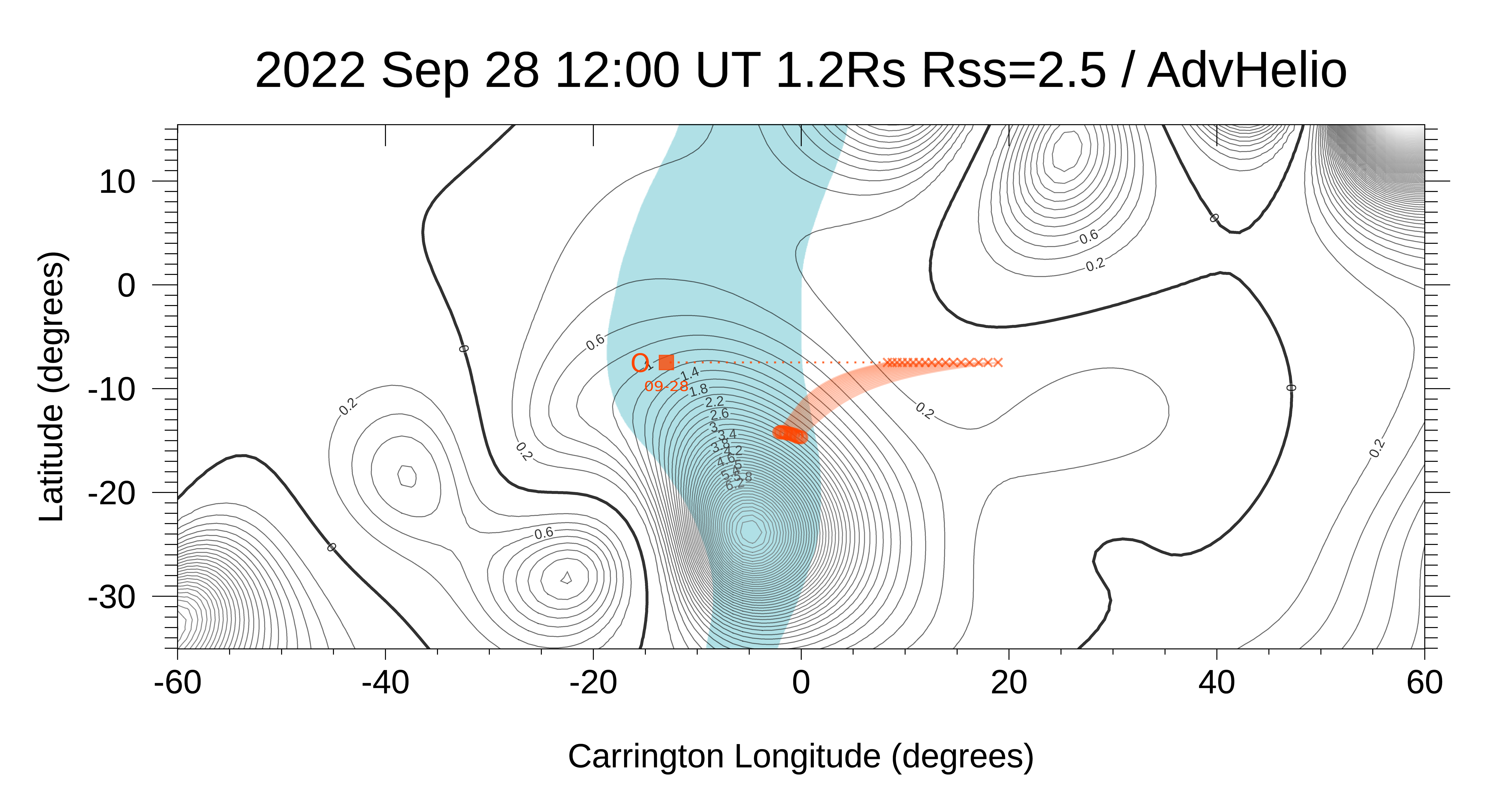}
    \includegraphics[width=11cm]{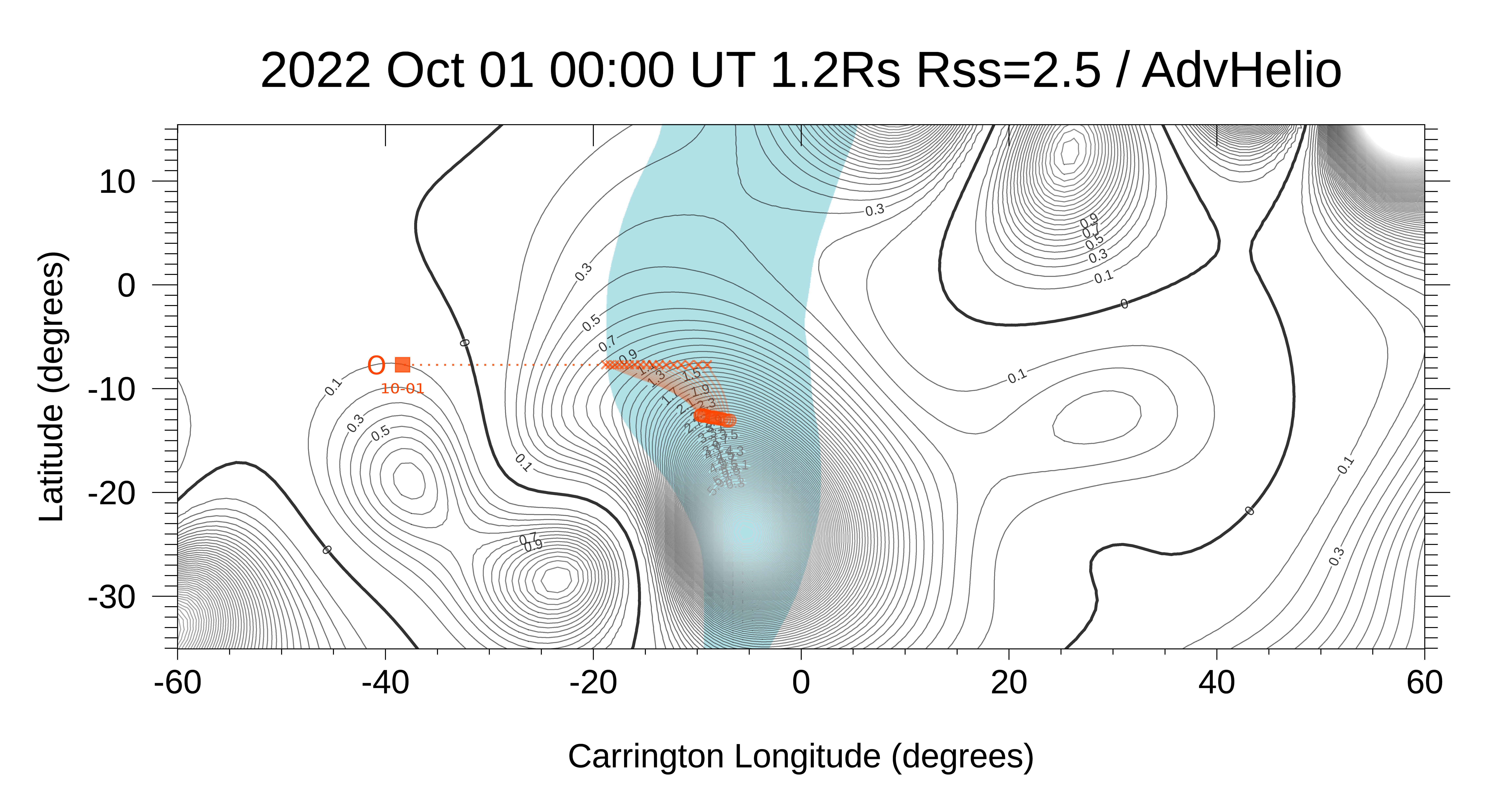}
    \includegraphics[width=11cm]{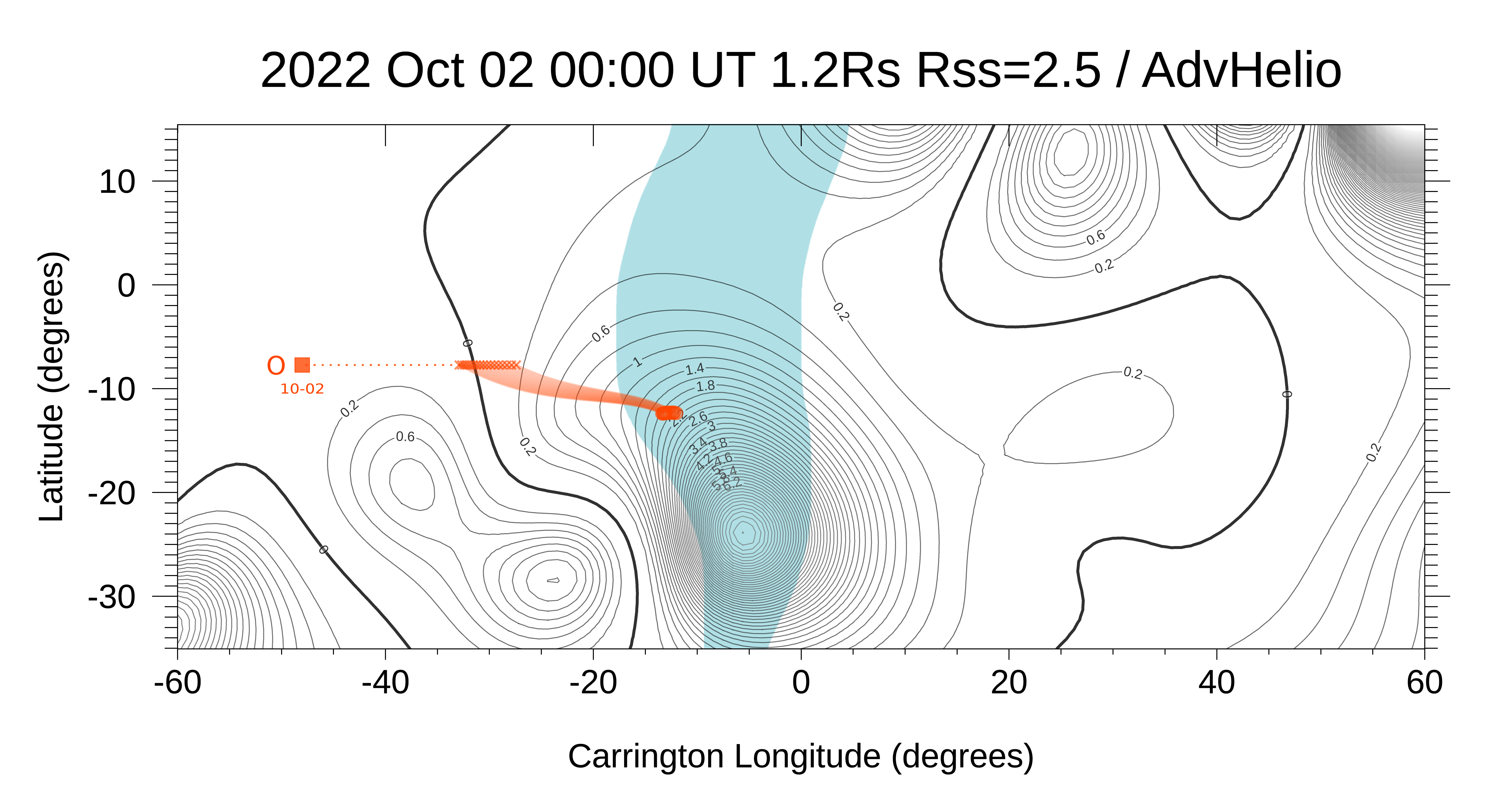}
      \includegraphics[width=0.4\linewidth]{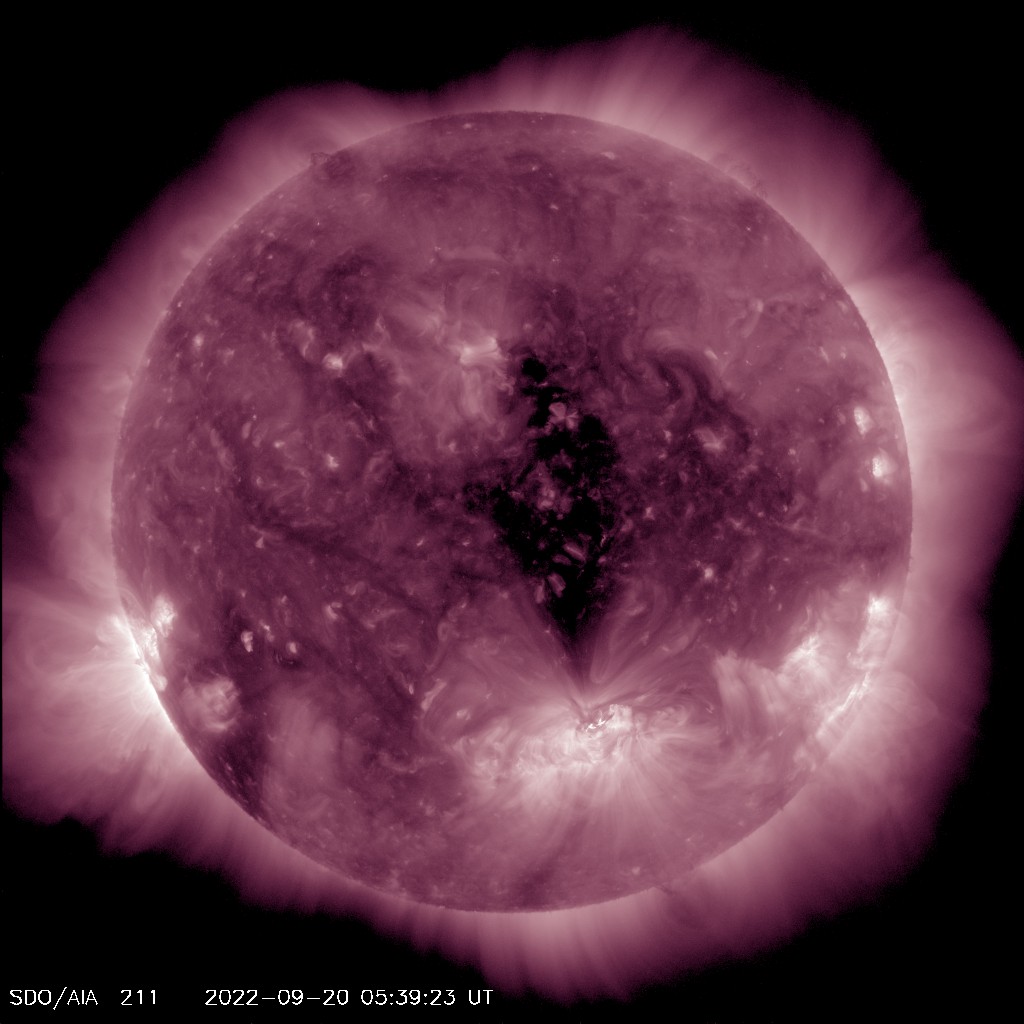}
    \caption{Upper three panels: Potential field source surface $B^2$ contour maps and solar wind magnetic foot-points along Solar Orbiter  trajectories (red) for the selected intervals. The maps show magnetic pressure iso-contours calculated for the heights R = 1.2 R$_{\odot}$ and the projection of the s/c location (squares) on the source surface (crosses) and down to the solar wind source region (circles). The crosses result from ballistic mapping using the measured in situ solar wind speed $\pm80$ $km~s^{-1}$ in bins of 10 $km~s^{-1}$. Open magnetic field regions are shown in cyan (negative polarity) while the neutral line is in black bold. Bottom plot: Image from SDO/AIA of the 211$\AA$ band showing the corona and source regions of solar wind streams AS2 AS3 and FH described in the paper. The image date was chosen to best display the source region on the Sun.}
    \label{fig05c}
\end{figure*}

The magnetic connectivity of the Alfvénic streams to their source regions was investigated using the potential field source surface (PFSS) extrapolation \citep{DeRosa2003}. In this model, the magnetic field is assumed to be potential from the photosphere out to a spherical surface, located for the present calculations at R = 2.5 R$_{\odot}$ from the Sun center, where the field is assumed to become radial. PFSS extrapolation allows for assimilation of data from the Solar Dynamic Observatory (SDO) Helioseismic and Magnetic Imager (HMI) with a 6 hour cadence and includes a flux-dispersal model, evolving the field over the full solar photosphere. The PFSS model is used in combination with a ballistic projection backwards from the spacecraft location using the solar wind speed measured at Solar Orbiter. To correct for solar wind acceleration and test the stability of the projection, the ballistic projection is also carried out adding speeds of up to $\pm$ 80 km s$^{-1}$ in bins of 10 km s$^{-1}$ \citep[e.g., see][and references therein]{Pan20}. The upper panel of Figure \ref{fig05a}, the three upper panels of Figure \ref{fig05b} and the upper panel of Figure \ref{fig05c} respectively show the potential field source surface B$^2$ contour maps and solar wind magnetic foot-points along Solar Orbiter trajectories (red) for the selected intervals. The projected position of Solar Orbiter onto the Sun is indicated by 'O', while the crosses indicate the result of the ballistic projection with the added speeds and the circles then indicate the mapping back from the source surface to the Solar corona at a height of 1.2 $R_s$. 
The source region of the selected streams is illustrated using images taken by the SDO Atmospheric Imaging Assembly (AIA) in the 211 $\AA$ passband (bottom panel of Figure \ref{fig05a}, bottom right panel of Figure \ref{fig05b}, bottom panel of Figure \ref{fig05c}.  

Figure \ref{fig05a} relates to the first fast interval labeled 'F' (see Figure \ref{fig01}). In this period, Solar Orbiter was connected to the large coronal hole with wide southward extension.
Figure \ref{fig05b} relates to interval AS1, when Solar Orbiter traced back to an open field region in the neighborhood of a large pseudo-streamer, as seen by the green (open) and white (closed) field lines in the lower left panel. Note that the lower panel image shows the pseudostreamer directly facing the observer. The high, non-monotonic expansion factors associated with the pseudostreamer open fields are most probably the reason for the slower wind speeds. The 3D magnetic field configuration of the pseudostreamer for Sep 25$\rm ^{th}$ is presented in the middle panel of Figure \ref{fig05b}, highlighting its non-monotonic expansion branch.
Finally, Figure \ref{fig05c} covers the interval September 28$\rm ^{th}$ - October 2$\rm ^{nd}$, when the Solar Orbiter connection moves from through a negative polarity coronal hole, crossing a pseudo-streamer (AS2, AS3) that is then destroyed (faster stream on October 2$\rm ^{nd}$).  
These results are in agreement with previous studies that focused on the solar sources of Alfvénic streams  \citep{Bale2019,Pan20,damicis2021solo,ervin2024,damicis2025}.

\section{Discussion}
The maximum solar wind speed of plasma from typically slow wind sources is faster than the minimum solar wind speed from CHs.
The Alfvénic slow wind naturally shows up as the low speed extension of Alfvénic, helium-rich wind. This overlap in speeds is likely due to how different in transit acceleration of the solar wind born in typical fast and slow regions \citep{alterman2025}, implying that the energy density of Alfvénic fluctuations in the ASW is lower than that in solar wind that achieves typical fast wind speeds $\sim 600 \, \mathrm{km \; s^{-1}}$.

This study is framed within this context focusing on observations performed in the second half of September 2022 when all SWA sensors detected several Alfvénic streams. The different streams were selected on the basis of a high $C_{VB}$ correlation coefficient, which is a proxy for the Alfvénic content of the fluctuations, along with almost incompressible conditions and large amplitude fluctuations. 
In the selected intervals, magnetic fluctuations also show the presence of switchbacks, which are one manifestation of Alfvénic fluctuations.
We characterized and compared Alfvénic intervals with different bulk speeds, paying particular attention to the main solar wind constituents, and performed a comparative study between a fast wind, three Alfvénic slow wind intervals and an intermediate speed stream. 



For the selected intervals, SWA-PAS VDFs are generally anisotropic and show a prominent beam component, accounting for a different fraction of the main proton population according to the different stream we are observing, that spans from about 5\% to 30\%, in agreement with previous findings \citep{neugebauer1962,neugebauer1966,asbridge1974,yermolaev1997,kasper2007,brunodemarco}. The beam is located at about 1.6 - 2 times the Alfvén velocity as already found in previous studies in the inner heliosphere \citep{feldman1973,Marsch1982b,marschlivi1987,Goldstein2000,brunodemarco,damicis2025} and faster than typical beam velocities observed at 1 au \citep{alterman2018}. SWA-PAS $V_{\parallel}$ profiles can be fitted by a skewed distribution. The Normal Inverse Gaussian (NIG) formulation quantifies the heat flux associated to each stream \citep{louarn_24}, which is relevant to the study of acceleration and heating processes.

SWA-EAS observations mainly focus on the behavior of strahl electrons that are highlighted in 2D Pitch-angle distributions.  The comparison between fast and non-Alfvénic slow solar wind streams presented in our study is consistent with earlier studies \citep{pilipp1987a,Owen_2022}, which report that electron strahl populations are typically more highly collimated and exhibit narrower pitch-angle widths in fast wind, while becoming progressively broader—and in some cases nearly isotropic—under standard slow wind conditions. This general trend also aligns with kinetic-theory predictions \citep{Horaites2018} and with radial evolution studies based on Helios measurements \citep{Maksimovic2005}, both of which show that strahl broadening is enhanced in higher-density, more collisional, and more turbulent solar wind. In our study, we find that collimated strahl beams appear in both fast and Alfvénic slow wind streams; however, the beams observed in Alfvénic slow wind are systematically broader and less intense than those detected in the fast wind. This suggests that electrons in Alfvénic slow wind may have undergone stronger or longer-duration pitch-angle scattering during their outward propagation. Such scattering may plausibly be associated not only with Coulomb collisions but also with wave–particle interactions, such as whistler-mode heat-flux instabilities \citep{Roberg2018,Lopez2019}, which have been proposed as an efficient mechanism for regulating the heat flux and diffusing strahl electrons into the halo population. This observational evidence implies that the electron populations within the highlighted sub-intervals are capable of transporting a larger heat flux than those in the surrounding non-Alfvénic slow wind, reflecting significant differences in the efficiency of scattering and in the overall evolution of the suprathermal electron distribution within different solar wind regimes.

Analysis of SWA-HIS heavy ion composition confirms differences in the solar wind streams observed during the study period. The O and C charge-state ratios were significantly lower during the Fast interval, characteristic of the coronal hole-associated (fast) solar wind. In contrast, the AS1, AS2, and AS3 intervals exhibit higher O and C ratios more typical of non-coronal hole-related (slow) wind, despite not always exceeding the conventional slow wind thresholds \citep{zhao2009}. The FH interval presents intermediate values of the charge state ratios. Correlation analysis shows that the Fast interval is distinctly separated from the AS and FH intervals in the composition space and none of the intervals fall into the outlier wind regime \citep{zhao2017}. Compared to earlier results from 2002 \citep{damicis2019}, the 2022 Fast wind displays much lower O and C ratios, more consistent with solar minimum values \citep{lepri2013}, possibly reflecting solar-cycle variability of composition signatures. Furthermore, AS1-3 and FH wind show intermediate composition between that of typical slow and fast wind values, supporting recent findings linking these intervals to solar wind originating near over-expanded coronal hole boundaries \citep{ervin2024}.

The investigation of the magnetic connectivity shows that the fast wind is connected to a large coronal hole with a wide southward extension while AS1 traced back to an open field region in the neighborhood of a large pseudostreamer, characterized by a strong expansion factor \citep{Pan20}. The non-monotonic expansion factors associated with the pseudostreamer open fields are most probably the reason for the slower wind speeds observed during the last part of the stream. On the other hand, AS2, AS3 and FH show a connection that moves from through a negative polarity coronal hole crossing a pseudostreamer (AS2, AS3). The latter is then destroyed, explaining the presence of a faster solar wind on October 1st. These findings are consistent with previous studies mapping back the Alfvénic slow wind to over-expanded open field source regions \citep{Pan20,damicis2021solo,damicis2025}.

One of the aims of the present study is to characterize and compare Alfvénic fluctuations in the different streams. We first addressed this by comparing velocity fluctuations with the corresponding magnetic field fluctuations in Alfvén units. The scatter plot of their $N$ components was used to derive a first approximation of the level of correlation, as well as the energy balance of the fluctuations by computing $r_A$ from the slope of the scatter plot. The derived values of $r_A$ values show energy equipartition in the fast wind, a clear magnetic energy imbalance in AS2, and only a slight imbalance in the remaining intervals. 
A more complete characterization of Alfvénic fluctuations is provided by spectral analysis. To this end, we investigated $\sigma_C(f)$ and $\sigma_R(f)$ in the frequency domain to identify the Alfvénic range. Despite limitations imposed by instrumental noise—primarily affecting plasma measurements at high frequencies \citep[see also][for a detailed discussion]{damicis2025}—the comparative spectral analysis reveals similar extensions of the Alfvénic range across the different streams. These findings are consistent with previous Solar Orbiter and Helios observations \citep{tumarsch1995,damicis2022,damicis2025}. Moreover, the higher temporal resolution of SWA–PAS enables the exploration of higher frequencies, corresponding to timescales of the order of minutes, which were not accessible to Helios.
The tendency of $\sigma_C$ to approach zero at high frequencies is associated with a flattening of the $e^-$ spectrum rather than a decrease of $e^+$, in agreement with earlier studies \citep{damicis2022, damicis2025}. This behavior has been interpreted as a consequence of the parametric decay of outward-propagating Alfvénic fluctuations at high frequencies \citep{tu1989, galeev1963}, invoked to explain the observed flattening of the $e^-$ spectrum. Alternatively, inward fluctuations may have a compressive nature \citep{MarschTu1990b}, possibly related to magnetoacoustic waves, pressure-balanced structures \citep{Goldstein1972}, or fine-scale stream tubes \citep{Thieme1987}. \citet{Bruno1989} suggested that the decrease in cross-helicity may be accompanied by the emergence of strong compressive magnetic fluctuations rather than by the local generation of inward-propagating Alfvénic modes. Within this framework, the $e^-$ spectrum could result from magnetic field directional turnings (MFDTs), assuming an interplanetary turbulence dominated by outward-propagating Alfvén waves and convected structures \citep{tumarsch1991}.
Although a strong correlation between inward-oriented Alfvénic fluctuations and relative density fluctuations has been reported \citep{Grappin1990}, caution is required when interpreting such signatures. Indeed, \citet{demarco2020} demonstrated that instrumental effects may introduce artificial spectral and kinetic features that can mimic physical processes.
Finally, it is worth noting that all Alfvénic streams, with the exception of AS2, remain close to energy equipartition despite their different heliocentric distances (0.58–0.38 au) and solar wind bulk speeds. More specifically, AS1, AS3, and FH show a slight magnetic energy excess at low frequencies (below $10^{-3}$ Hz) and energy equipartition in the range $10^{-3}$–-$2 \times 10^{-2}$ Hz. At higher frequencies, the spectra are dominated by instrumental noise, preventing any firm conclusions.
For completeness, it is worth recalling that the departure from energy equipartition has also been attributed to turbulence evolution \citep[e.g.,][]{grappin, roberts1992} and to the presence of convected solar wind structures \citep{tumarsch1993}.

Another aspect to consider is the expected radial evolution of the turbulent behavior of solar wind fluctuations \citep[e.g.,][]{tumarsch1995,brunocarbone2013,Telloni2015}. Indeed, by combining the results of \citet{damicis2022} at 1 AU and \citet{damicis2025} at $\sim$0.32 AU, a clear radial evolution of turbulence emerges.
In the present work, however, although the spacecraft undergoes a radial excursion of about 0.2 au—which should, in principle, allow us to detect some degree of radial evolution—such effects are not observed in our data. In particular, one would expect a shift of the Alfvénic range toward lower frequencies (i.e., larger timescales), together with a decrease in the magnitude of $\sigma_R$ and a tendency for $r_A$ to become smaller than unity \citep{marschtu1990}. Although $\sigma_R$ is known to be close to zero near the Sun, it is expected to decrease significantly with increasing heliocentric distance \citep[e.g.,][]{bruno1985,bruno2007, marschtu1990,damicis2022}.

Contrary to these expectations, our results show that solar wind fluctuations remain close to energy equipartition in the fast wind far from perihelion, as well as in some Alfvénic slow wind intervals. These latter streams would be expected to evolve more rapidly and depart from equipartition. Remarkably, AS2 and AS3—both Alfvénic slow wind intervals observed at nearly the same heliocentric distance and originating from the same solar source—exhibit different energy contents in their fluctuations. Such differences in energy balance may be related to distinct processes acting during solar wind acceleration and interplanetary transport \citep{Halekas2023, Rivera2024, alterman2025}. A detailed study combining in situ turbulence parameters with the helium abundance \citep{alterman2025a,alterman2025d} may confirm our mappings of the observed solar wind streams to their source regions. These intriguing results deserve further investigation and will be addressed in future work.
Taken together, these results suggest that the variability observed among the different streams cannot be simply ascribed to radial evolution effects. Instead, significant differences persist even among streams observed at comparable heliocentric distances and originating from similar coronal source regions, pointing to the role of additional physical processes acting either at the source or during the evolution in the interplanetary medium.

The comparison among the different streams highlights several differences. The fast wind is characterized by large-amplitude fluctuations and by magnetic and velocity fluctuations close to energy equipartition, and it originates from a large low-latitude coronal hole. A similar behavior is observed for AS2, AS3, and FH, all of which map back to the same open-field region, namely a negative-polarity coronal hole extending over several tens of degrees in latitude.
Remarkably, AS2 exhibits a magnetic energy imbalance, at odds with the other intervals, despite originating from the same source region. In contrast, AS1 is characterized by smaller relative fluctuation amplitudes compared to the other streams and, unlike them, originates from a large pseudostreamer.

\section{Conclusion}

In a broad, statistical sense, the Alfvénic slow wind is emerging as a third class of solar wind with speeds typical of the nominal slow wind and with other properties typical of the fast wind.
These differences likely reflect variations in the magnetic topology of the different solar wind source regions. Such variations would impact both the processes that determine solar-wind charge-state and composition ratios \citep{vonsteiger2000,geiss1995,geiss1995b,Xu2015,lepri2013,zhao2022,alterman2025b,Shi2025} and the Alfvénic content of the solar wind, which in turn affects its transit acceleration profiles \citep{alterman2025,Halekas2023,Rivera2024,Rivera2025}.

We have analyzed three slow, Alfvénic solar wind streams along with a fast wind stream and a moderate fast stream observed at a heliocentric distance below 0.6 au by Solar Orbiter from September 14th to October 2nd, 2022 covering a distance of 0.2 au.
We have compared the proton VDFs, electron VDFs, charge state ratios, cross helicity, residual energy, and normalized power spectra and studied the magnetic connectivity.
The streams are defined as Alfvénic based on the classical definition of 'Alfvénicity'. 
We observe proton beams in the VDFs of all the selected streams, along with pronounced electron strahl populations that are narrower and more intense in the fast and FH intervals compared to the other AS regions. The presence of beams in both proton and electron VDFs is typical of fast and AS streams.
Similarly, the charge state ratios are significantly lower during the fast wind, which is typical of coronal hole-associated solar wind, compared to the AS intervals, which exhibit ratios more typical of non-coronal hole-related (slow) wind, and the FH showing intermediate values.
Thus, these observations support the view that the ASW is the slow speed extension of fast wind.
On the other hand, the present study shows that solar wind fluctuations are close to energy equipartition in the fast wind and for some Alfvénic slow wind intervals, while the AS2 shows magnetic energy imbalance. It is also remarkable that AS2 and AS3, both Alfvénic slow wind intervals, observed at almost the same heliocentric distance and coming from the same solar source, show a different energy content of the fluctuations. This substantiates near-Sun observations of varying magnetic field fluctuation amplitude within streams from the same source region \citep{Rivera2024b}.
Further analysis combining solar wind helium abundance and cross helicity with solar wind speed \citep{alterman2025,alterman2025a,alterman2025b,alterman2025c,alterman2025d} should be applied to these and similar streams to gain additional insight. 

These findings also present a wide range of challenges to existing models of the solar wind, particularly in explaining how Alfvénicity can be maintained within slow solar wind streams that originate from distinct regions on the Sun. The stationarity or evolution of the solar source could partly explain why Alfvénicity can be maintained or be lost within slow solar wind streams when observed at different heliocentric distances \citep[e.g.,][]{damicis2021rev}. However, one should also take into account local inhomogeneities that would determine the degradation of v-b correlations as, for example, the presence of compressible phenomena \citep[e.g.,][]{BB1991}.
In particular, it is unclear why, in some regions, the solar wind remains highly Alfvénic as it propagates away from the Sun. On the other hand, in which conditions and why some Alfvénic streams depart from energy equipartition and thus show a different behavior with respect to previous cases \citep[e.g.,][]{tumarsch1995,damicis2022,damicis2025} is still not understood and needs further investigation. This raises fundamental questions about the interplay between magnetic field topology, wave-particle interactions, coronal heating processes, and the role of turbulence and compressions in shaping solar wind properties. 
The fact that Alfvénic fluctuations persist across the heliosphere suggests the presence of robust and possibly universal mechanisms operating in the early phases of solar wind formation and acceleration. High-resolution measurements from Solar Orbiter, combined with those from Parker Solar Probe, will provide unprecedented access to the inner heliosphere, allowing scientists to probe the nascent solar wind closer to its source than ever before with particular reference to the Alfvénic solar wind in a broad range of heliocentric distances and temporal scales. 
Despite these observational advances, more statistics is needed and the combined observations from all the SWA sensors are crucial to study the underlying physical processes that give rise to the variety of outcomes observed in solar wind streams that remain still not fully understood.

\begin{acknowledgements}
Solar Orbiter is a mission of international cooperation between ESA and
NASA, operated by ESA. Solar Orbiter SWA data were derived from
scientific sensors that were designed and created and are operated under
funding provided by numerous contracts from UKSA, STFC, the Italian
Space Agency, CNES, the French National Centre for Scientiﬁc
Research, the Czech contribution to the ESA PRODEX
program, and NASA. Solar Orbiter SWA work at INAF/IAPS is currently funded under ASI grant 2018-30-HH.1-2022.
Solar Orbiter SWA work at the UCL/Mullard Space Science Laboratory is currently funded by STFC (grant Nos. ST/W001004/1 and ST/X/002152/1).

This research was supported by the International Space Science Institute (ISSI) in Bern through the ISSI International Team projects
\#550 ``Solar Sources and Evolution of the Alfvénic Slow Wind'', \#560 ``Turbulence at the Edge of the Solar Corona: Constraining Available Theories Using the Latest Parker Solar Probe Measurements'', and \#23-591 ``Evolution of Turbulence in the Expanding Solar Wind''.
LSV was supported by the Swedish Research Council (VR) Research Grant N. 2022-03352, and by the projects “2022KL38BK– The ULtimate fate of TuRbulence from space to laboratory plAsmas (ULTRA)” (B53D23004850006) and ‘Data-based predictions of solar energetic particle arrival to the Earth: ensuring space data and technology integrity from hazardous solar activity events’ (H53D23011020001), funded by the Italian Ministry of University and Research through PRIN/NRRP-NextGenerationEU.

\end{acknowledgements}

%
%

\begin{appendix}
\section{Tools for studying turbulence}
\label{A}

Els\"asser variables are defined in the following way: $z^{\pm} = v \pm b$, where $v$ and $b$ are the velocity and magnetic fluctuating field where the average background filed as been substracted. In particular, $b$ is the magnetic field vector in Alfvén units for a background magnetic field pointing towards the Sun, while $z^{\pm} = v \mp b$ for the opposite polarity. This definition allows to always identify $z^+$ with outward propagation modes (with respect to the Sun) while $z^-$ with inward modes (fluctuations).

In particular, we computed the three components of $z^+$ and $z^-$ in RTN and then the PSD of the trace of their components, $e^{\pm}$ vs $f$, that is the energy associated to these propagating modes, either inward or outward. 
We then derived the normalized cross-helicity in the frequency domain: $\sigma_C (f) = (e^+(f) - e^-(f))/(e^+(f) + e^-(f))$, that measures the predominance of $z^+$ on $z^-$ or vice-versa. $\sigma_C$ equal to 1 (-1) indicates the presence of only the outward (inward) component, while $\vert \sigma_C \vert < 1$ corresponds to mixed inward and outward modes and/or non-Alfvénic fluctuations \citep{brunocarbone2013}.

The energy imbalance of the fluctuations in the spectral domain can be evaluated by computing first the trace of the power spectral density of velocity components (here indicated as $e^V(f)$), and $b$ in Alfvén units, $e^b(f)$. The ratio between these two quantities corresponds to the Alfvén ratio. We then define the normalized residual energy in the frequency domain, $\sigma_R(f) = (e^V(f) - e^b(f))/(e^V(f) + e^b(f))$. $\sigma_R$ is linked to $r_A$ by $(r_A - 1)/(r_A + 1)$. $\sigma_R$ equal to $+1$ ($-1$) indicates the absence of magnetic (kinetic) fluctuations. Pure Alfvénic fluctuations are expected to show equipartition of energy: $\sigma_R = 0$ (or $r_A = 1$) \citep{tumarsch1995}.
    
\section{Role of the different ion populations in the energy balance of solar wind fluctuations}
\label{B}

In this section, we further investigate the tendency toward a kinetic energy imbalance observed in the fast wind (Figure~\ref{fig07}) by exploiting the clustering technique originally introduced by \citet{demarco2023} and subsequently developed by \citet{brunodemarco}. This approach allows us to account for the contribution of alpha particles to the energetics of fluctuations \citep{bav2000}. In particular, \citet{brunodemarco} analyzed the kinetic properties of protons and alpha particles—distinguishing between their core and beam populations—in the same fast stream considered in the present study. Here, we use the same dataset to assess the role of the different ion populations and to determine whether the observed energy imbalance has a physical origin.

\begin{figure}
    \centering
    \includegraphics[width=0.8\columnwidth]{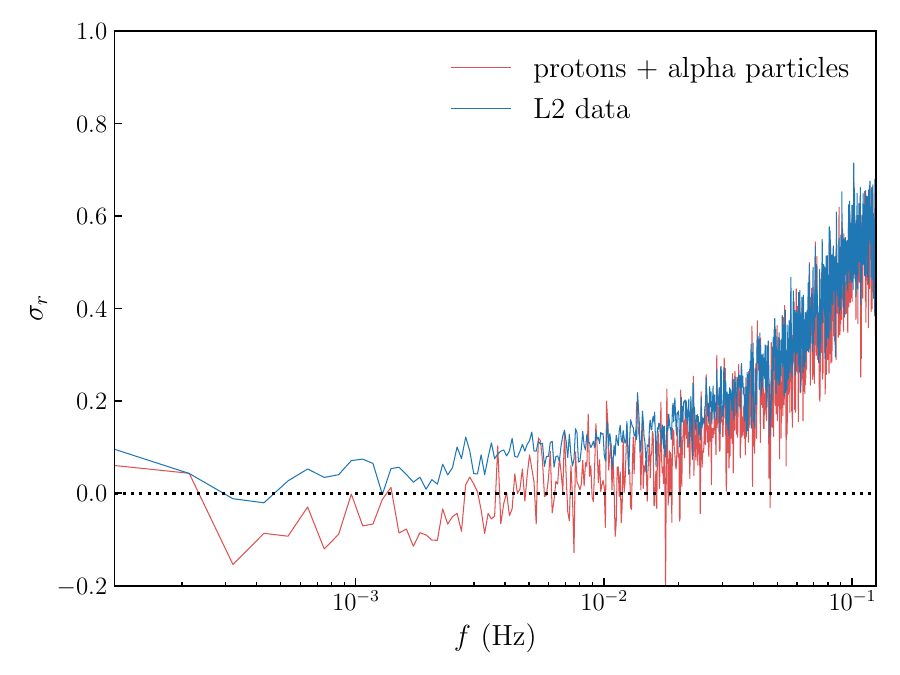}
    \caption{Comparison of the power spectra of the normalized residual energy, $\sigma_R$, for the fast wind, when considering the contribution of the different ion populations: protons + alpha particles (red) and L2 data (blue).}
    \label{fig_sigR}
\end{figure}

Clustering applied to 3D VDFs observed by SWA-PAS allows to separate the full 3D VDF in 4 3D VDFs (referred to proton core and beam, alpha core and beam) and then derives the moments (or bulk parameters) of distributions of the different ion species. 
Although \cite{brunodemarco} did not perform a spectral analysis, Figure 21 of their paper already gave an indication of a kinetic imbalance for the proton core since they found a slope in the scatter plot of $V_N$ vs $V_{AN}$ close to 1.5. In this sense, the balance of energy expected for ideal alfvénic fluctuations can be properly reached when all the populations are included in the calculations. This argument can explain the findings of Figure \ref{fig07} of the present paper (fast wind, middle panel), where we show a $\sigma_R$ slightly imbalanced towards positive values. Indeed, by separating the population of the fast wind interval using the clustering technique, we performed the spectral analysis of $\sigma_R$ by considering the velocity obtained from protons and alpha particles. In this case, the mass density used to define $V_A$ takes into account also the contribution of both protons and alpha particles.
Figure \ref{fig_sigR} shows the comparison of the spectrum calculated by including the contribution of proton and alpha particles along with the $\sigma_R$ spectrum obtained by using L2 data.
From this figure, it is clear that accounting for all the contributions from the different solar wind populations is essential to correctly assess the energetics of fluctuations, in agreement with the model proposed by \citet{brunodemarco}. Through the present spectral analysis, this concept is further elucidated in the frequency domain. In particular, while moments computed using proton-only measurements (L2 data) indicate a slightly positive energy imbalance, the inclusion of alpha particles in the calculation leads the residual energy back toward equipartition.

\end{appendix}

\end{document}